\documentclass[sigconf]{acmart}
\pdfoutput=1

\AtBeginDocument{%
  \providecommand\BibTeX{{%
    \normalfont B\kern-0.5em{\scshape i\kern-0.25em b}\kern-0.8em\TeX}}}

\setcopyright{acmcopyright}
\copyrightyear{2022}
\acmYear{2022}
\setcopyright{acmcopyright}\acmConference[MM '22]{Proceedings of the 30th ACM International Conference on Multimedia}{October 10--14, 2022}{Lisboa, Portugal} \acmBooktitle{Proceedings of the 30th ACM International Conference on Multimedia (MM '22), October 10--14, 2022, Lisboa, Portugal}
\acmPrice{15.00}
\acmDOI{10.1145/3503161.3548072}
\acmISBN{978-1-4503-9203-7/22/10}

\usepackage[linesnumbered,boxed,ruled]{algorithm2e}
\usepackage{enumitem}
\usepackage{dsfont}
\usepackage{graphicx}

\usepackage{multirow}

\usepackage{subfigure}

\usepackage{ragged2e}
\usepackage{endnotes}

\newcommand{\nosection}[1]{\vspace{2pt}\noindent\textbf{#1.}}

\newcommand{\modelname}{\textbf{DDGHM}}

\begin{document}

\title{\modelname: Dual Dynamic Graph with Hybrid Metric Training for Cross-Domain Sequential Recommendation}

\author{Xiaolin Zheng}
\affiliation{
\institution{College of Computer Science, Zhejiang University}
\city{Hangzhou}
\country{China}
}
\email{xlzheng@zju.edu.cn}

\author{Jiajie Su}
\affiliation{
\institution{College of Computer Science, Zhejiang University}
\city{Hangzhou}
\country{China}
}
\email{sujiajie@zju.edu.cn}

\author{Weiming Liu}
\affiliation{
\institution{College of Computer Science, Zhejiang University}
\city{Hangzhou}
\country{China}
}
\email{21831010@zju.edu.cn}

\author{Chaochao Chen}
\authornote{Chaochao Chen is the corresponding author.}
\affiliation{
\institution{College of Computer Science, Zhejiang University}
\city{Hangzhou}
\country{China}
}
\email{zjuccc@zju.edu.cn}


\renewcommand{\shortauthors}{Xiaolin Zheng, Jiajie Su, Weiming Liu, \& Chaochao Chen}

\ccsdesc[500]{Information systems~Recommender systems}

\begin{abstract}

Sequential Recommendation (SR) characterizes evolving patterns of user behaviors by modeling how users transit among items.
%
However, the short interaction sequences limit the performance of existing SR. 
To solve this problem, we focus on Cross-Domain Sequential Recommendation (CDSR) in this paper, which aims to leverage information from other domains to improve the sequential recommendation performance of a single domain. 
%
Solving CDSR is challenging. 
On the one hand, how to retain single domain preferences as well as integrate cross-domain influence remains an essential problem.
On the other hand, the data sparsity problem cannot be totally solved by simply utilizing knowledge from other domains, due to the limited length of the merged sequences. 
To address the challenges, we propose \modelname, a novel framework for the CDSR problem, which includes two main modules, i.e., dual dynamic graph modeling and hybrid metric training. 
The former captures intra-domain and inter-domain sequential transitions through dynamically constructing two-level graphs, i.e., the local graphs and the global graphs, and incorporating them with a fuse attentive gating mechanism. 
The latter enhances user and item representations by employing hybrid metric learning, including collaborative metric for achieving alignment and contrastive metric for preserving uniformity, to further alleviate data sparsity issue and improve prediction accuracy. 
We conduct experiments on two benchmark datasets and the results demonstrate the effectiveness of \modelname.

\end{abstract}

\keywords{Cross-Domain Recommendation; Sequential Recommendation; Dynamic Graph}

\maketitle

\setlength{\floatsep}{4pt plus 4pt minus 1pt}
\setlength{\textfloatsep}{4pt plus 2pt minus 2pt}
\setlength{\intextsep}{4pt plus 2pt minus 2pt}
\setlength{\dbltextfloatsep}{3pt plus 2pt minus 1pt}
\setlength{\dblfloatsep}{3pt plus 2pt minus 1pt}
\setlength{\abovecaptionskip}{3pt}
\setlength{\belowcaptionskip}{2pt}
\setlength{\abovedisplayskip}{2pt plus 1pt minus 1pt}
\setlength{\belowdisplayskip}{2pt plus 1pt minus 1pt}

\section{Introduction}


Sequential Recommendation (SR) has attracted increasing attention due to its significant practical impact. 
SR aims to find the potential patterns and dependencies of items in a sequence, and understand a user’s time-varying interests to make next-item recommendation \cite{hidasi2015session,kang2018self,tang2018personalized,sun2019bert4rec,wang2020next}. 
Though some SR models have been proposed, they face the difficulty in characterizing user preferences when behavior sequences are short, e.g., data sparsity \cite{li2019zero}. 
%
Therefore, it is necessary to utilize more information, e.g., side information or data from other domains, to mitigate the above problem.
%

\begin{figure}[t]
 \begin{center}
 \includegraphics[width=\columnwidth]{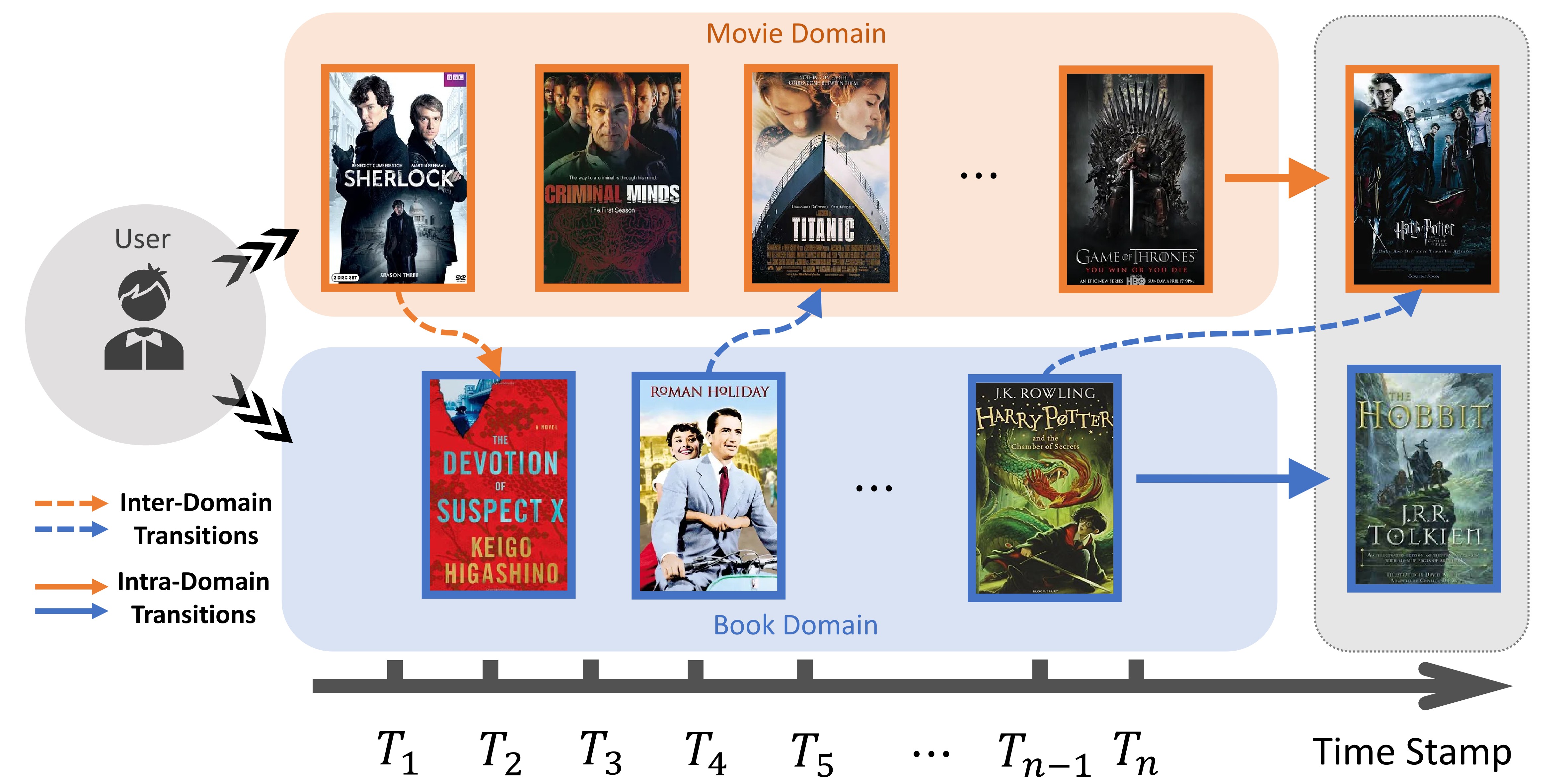}
 \vspace{-0.2 in}
 \caption{A motivating example. 
 }
 \vspace{-0.15 in} 
 \label{fig:story}
 \end{center}
\end{figure}

In this paper, we focus on the Cross-Domain Sequential Recommendation (CDSR) problem, which considers the next-item prediction task for a set of common users whose interaction histories are recorded in multiple domains during the same time period. 
Similar as how traditional Cross-Domain Recommendation (CDR) helps leverage information from a source domain to improve the recommendation performance of a target domain \cite{zhu2021cross}, CDSR shows its superiority by incorporating sequential information from different domains \cite{ma2019pi}. 
%
Intuitively, a user's preference could be reflected by his behaviors in multiple domains.
We motivate this through the example in Figure \ref{fig:story}, where a user has some alternating interactions in a movie domain and a book domain during a period of time. 
From it, we can easily observe that the user's choice for the next movie `Harry Potter' depends not only on his previous interest for mystery and fantasy movies (\textit{intra-domain}), but also on his reading experience of the original book `Harry Potter' (\textit{inter-domain}).

Solving the CDSR problem is challenging. 
On the one hand, complex sequential item transitions exist simultaneously inside domains and across domains, which makes it difficult to capture and transfer useful information.
On the other hand, although transferring auxiliary information from another domain helps explore sequential patterns, data sparsity problem still exists because a large number of items in both domains never or rarely appear in historical sequences.
Thus, how to effectively represent and aggregate both intra-domain and inter-domain sequential preferences as well as further alleviate data sparsity problem remains a crucial issue. 

%

Existing researches on SR, CDR, and CDSR cannot overcome these challenges well.
First, various methods have been proposed for SR \cite{rendle2010factorizing,hidasi2015session,tuan20173d,wu2019session,kang2018self,sun2019bert4rec}, but most of them only focus on the transition correlations in a single domain without considering the auxiliary information from other domains.
%
%
Second, although several representative CDR approaches \cite{hu2018conet,li2020ddtcdr,zhu2020graphical,chen2022differential} address the data sparsity and cold-start problem by transferring knowledge within domains, they cannot take full advantage of sequential patterns.
Third, existing CDSR methods \cite{ma2019pi,sun2021parallel,guo2021gcn,li2021dual,chen2021dual} reach the breakthrough of exploring sequential dependencies and modeling structure information that bridges two domains. 
%
%
%
%
%
However, these CDSR models cannot extract and incorporate both the intra-domain and the inter-domain item transitions in a dynamical and synchronous way.
%
%
Besides, the long-standing data sparsity problem is still overlooked.


To overcome the above limitations, we propose \modelname, a dual dynamic graphical model with hybrid metric training, to solve the CDSR problem.
%
The purpose of \modelname~is twofold: (1) exploit and fuse the evolving patterns of users’ historical records from two aspects, i.e., intra-domain and inter-domain.
(2) Enhance representation learning to address the remaining data sparsity problem so as to further improve sequential recommendation performance.
%
To do that, we build two modules in \modelname, i.e., \textit{dual dynamic graph modeling} and \textit{hybrid metric training}.
%
(i) The \textit{dual dynamic graph modeling module} simultaneously constructs separate graphs, i.e., local dynamic graphs and global dynamic graphs, to encode the complex transitions intra-domain and inter-domain respectively.
The structure of directed graph equipped with gated recurrent units makes it possible to integrate users' long-term and short-term interest discriminatively into the sequential representation.
Meanwhile, a novel gating mechanism with specialized fuse attention algorithm is adopted to effectively filter and transfer cross-domain information that might be useful for single domain recommendation. 
Dual dynamic graph modeling generates both item representations and complete sequence representations that indicate users' general preferences, and send them into (ii) the \textit{hybrid metric training module}. 
%
Hybrid metric training not only completes estimating recommendation scores for each item in both domains, but also alleviates the impact of data sparsity in CDSR by employing representation enhancement in two perspectives, i.e., a) \textit{collaborative metric} for realizing alignment within similar instances and b) \textit{contrastive metric} for preserving uniformity of feature distributions.
%

We summarize our main contributions as follows.
(1) We propose an innovative framework \modelname~that effectively leverages cross-domain information to promote recommendation performance under CDSR scenario.
(2) We design a dual dynamic graph modeling module which captures both intra-domain and inter-domain transition patterns and attentively integrate them in an interpretable way.
(3) We develop a hybrid metric training module which weakens the data spasity impact in CDSR by enhancing representation learning of user preferences.
(4) We conduct experiments on two real-word datasets and the results demonstrate the effectiveness of \modelname.

\section{Related Work}

\nosection{Sequential Recommendation}
Sequential Recommendation (SR) is built to characterize the evolving user patterns by modeling sequences. 
Early work on SR usually models the sequential patterns with the Markov Chain assumption \cite{rendle2010factorizing}. 
%
With the advance in neural networks, Recurrent Neural Networks (RNN) based \cite{hidasi2015session,hidasi2018recurrent,wu2017recurrent,donkers2017sequential}, Convolutional Neural Networks (CNN) based \cite{tang2018personalized}, Graph Neural Networks (GNN) based \cite{wu2019session,qiu2019rethinking,zheng2019balancing,zheng2020dgtn} methods, and Transformers \cite{kang2018self,sun2019bert4rec} have been adopted to model the dynamic preferences of users over behavior sequences.
%
%
%
%
%
Recently, unsupervised learning based models \cite{xie2020contrastive,qiu2021memory} are introduced to extract more meaningful user patterns by deriving self-supervision signals.
%
While these methods aim to improve the overall performance via representation learning for sequences, they suffer from weak prediction power for cold-start and data sparsity issues when the sequence length is limited.

\nosection{Cross-Domain Recommendation}
Cross-Domain Recommendation (CDR) is proposed to handle the long-standing cold-start and data sparsity problems that commonly exist in traditional single domain recommender systems \cite{li2009can,pan2013transfer}.
The basic assumption of CDR is that different behavioral patterns from multiple domains jointly characterize the way users interact with items \cite{zhu2019dtcdr}.
According to \cite{zhu2021cross}, existing CDR have three main types, i.e., \textit{transfer-based}, \textit{multitask-based}, and \textit{clustered-based} methods.
Transfer-based methods \cite{man2017cross,zhao2020catn} learn a linear or nonlinear mapping function across domains.
Multitask-based methods \cite{hu2018conet,zhu2020graphical,liu2022collaborative} enable dual knowledge transfer by introducing shared connection modules in neural networks.
Clustered-based methods \cite{wang2019solving} adopt co-clustering approach to learn cross-domain comprehensive embeddings by collectively leveraging single-domain and cross-domain information within a unified framework.
However, conventional CDR approaches cannot perfectly solve the CDSR problem, because they fail to capture sequential dependencies that commonly exist in transaction data.
%

\nosection{Cross-Domain Sequential Recommendation}
Existing researches on CDSR can be divided into three categories, i.e., RNN-based, GNN-based, and attentive learning based methods. 
%
%
First, RNN-based methods, e.g., $\pi$-Net \cite{ma2019pi} and PSJNet \cite{sun2021parallel}, employ RNN to generate user-specific representations, which emphasize exploring sequential dependencies but fail to depict transitions among associated entities.
Second, GNN-based method DA-GCN \cite{guo2021gcn} devises a domain-aware convolution network with attention mechanisms to learn node representations, which bridges two domains via knowledge transfer but cannot capture inter-domain transitions on the item level.
Third, attentive learning based methods \cite{li2021dual,chen2021dual} adopt dual attentive sequential learning to bidirectionally transfer user preferences between domain pairs, which show the effectiveness in leveraging auxiliary knowledge but cannot excavate structured patterns inside sequential transitions. 
%
%
%
%
%
To summary, the existing CDSR methods cannot extract and integrate both inter-domain and intra-domain information dynamically and expressively, and they all neglect the data sparsity issue which still remains after aggregating information from multiple domains.

\section{The Proposed Model}


\subsection{Problem Formulation}

CDSR aims at exploring a set of common users' sequential preference with given historical behavior sequences from multiple domains during the same time period.
Without loss of generality, we take two domains (A and B) as example, and formulate the CDSR problem as follows. 
We represent two single-domain behavior sequences of a user as $S_A =\{A_1,A_2,...,A_i,...\}$ and $S_ B=\{B_1,B_2,...,B_j,...\}$, where $A_i$ and $B_j$ are the consumed items in domain $A$ and $B$ respectively. 
Then the cross-domain behavior sequence $S$ is produced by merging $S_A$ and $S_B$ in the chronological order, for which the above example can be given as $S_M = \{B_1,A_1,B_2,...,A_i,B_j,...\}$.
Given $S_A$, $S_B$, and $S_M$, CDSR tries to predict the next item that will be consumed in domain $A$ and $B$.


\subsection{An overview of \modelname }

\begin{figure*}
\centering
\includegraphics[width=1\textwidth]{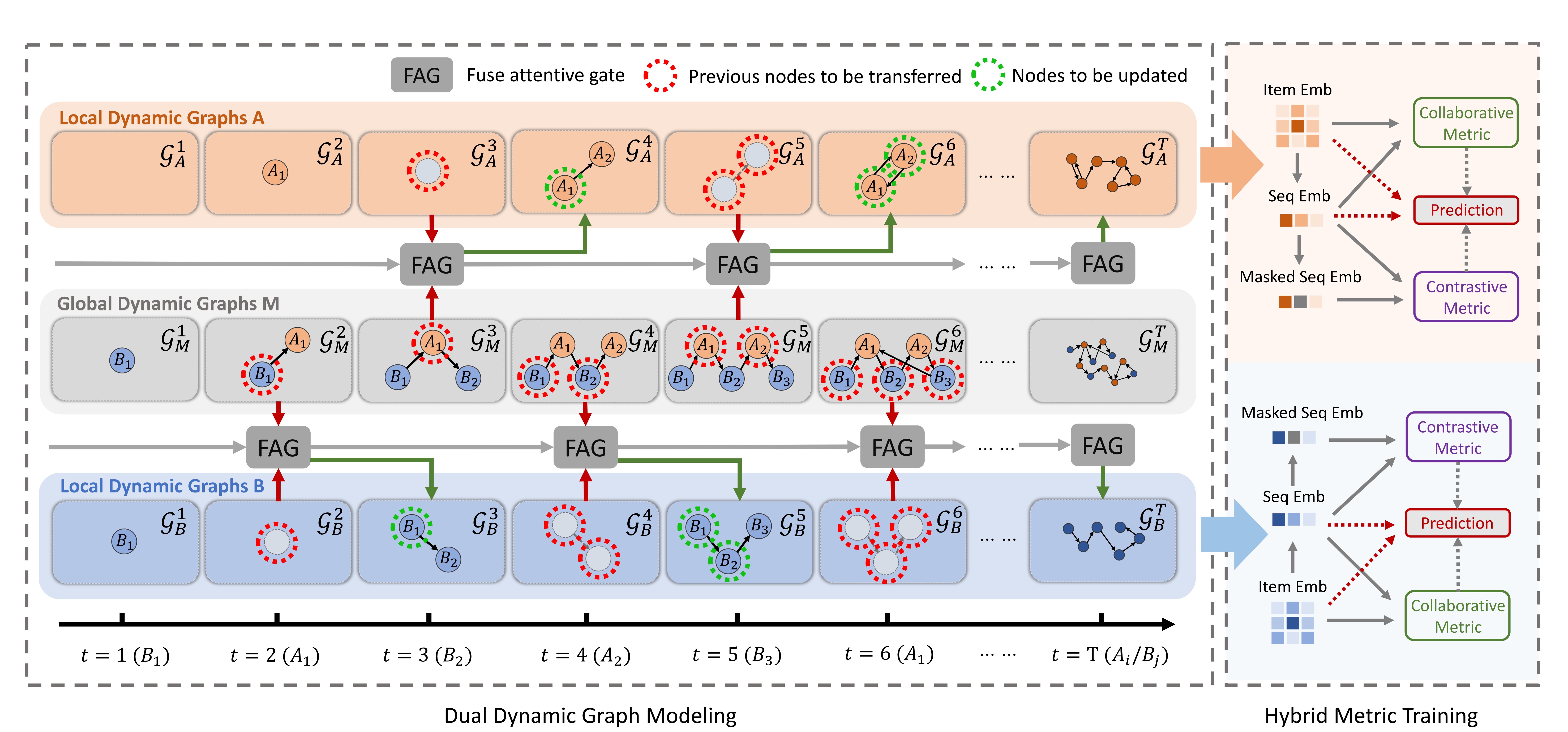}
\caption{An overview of \modelname~. Local dynamic graph for domain $A$ is shown in orange, domain $B$ is shown in blue, and global dynamic graph is shown in gray. Section 3.2 contains a walkthrough of the model. 
}
\label{fig:framework}
\end{figure*}

%
The aim of \modelname~is providing better sequential recommendation performance for a single domain by leveraging useful cross-domain knowledge.
The overall structure of our proposed \modelname~ is illustrated in Figure \ref{fig:framework}. 
%
%
%
\modelname~ consists of two main modules: (1) \textit{dual dynamic graph modeling} and (2) \textit{hybrid metric training}.
In the dual dynamic graph modeling module, we explore intra-domain and inter-domain preference features in parallel, and then transfer the information recurrently at each timestamp.
Thus, this module has two parts, i.e., \textit{dual dynamic graphs} and \textit{fuse attentive gate}.
In the dual dynamic graphs part, we build two-level directed graphs : a) \textit{local dynamic graphs} for extracting \textit{intra-domain} transitions; b) \textit{global dynamic graphs} for extracting \textit{inter-domain} transitions.
After it, we propose a fuse attentive gate which adopts attention mechanism to integrate item embeddings from global graphs into local graphs.
To this end, cross-domain information is effectively leveraged to enrich the single domain representations.
Later, sequence representations that combine intra-domain and inter-domain features are sent into the hybrid metric training module.

Though cross-domain information has been utilized, the data sparsity problem still exists.
Thus, in the hybrid metric training module, we propose the representation enhancement for optimizing representation learning so as to reduce data sparsity impact.
The enhancement is twofold : a) adopt collaborative metric learning to realize alignment between similar instance representations; b) employ contrastive metric learning to preserve uniformity within different representation distributions.  
Finally, the model outputs the probability of each item to be the next click for both domains.

\subsection{Dual Dynamic Graph Modeling}

The prime task for solving the CDSR problem is exploiting the expressive sequential representations to depict user interests from the cross-domain behavior sequences.
To this end, we propose a dual dynamic graph modeling module that extracts the latent item embeddings from both local and global dynamic graphs.

\subsubsection{Dual dynamic graphs}
We first introduce dual dynamic graphs, i.e., local dynamic graphs and global dynamic graphs. 
%

\nosection{Local dynamic graphs}
In this part, we apply a dynamic GNN to capture the intra-domain sequential transitions that represent user preference patterns in each single domain \cite{wu2019session, zhang2022dynamic}.
The graph modeling has two steps: \textbf{Step 1}: local dynamic graph construction, and \textbf{Step 2}: local dynamic graph representation.

\textit{\textbf{Step 1:} local dynamic graph construction.}
We first introduce how to convert the single-domain behavior sequence into a dynamic graph.
Taking domain $A$ as example,
%
%
%
a sequence can be represented by $S_A = [v_1,v_2,...,v_i,...]$, where $v_i$ represents a consumed item of the user within the sequence $S_A$.
The local dynamic graphs can be defined as $\{\mathcal{G}^1_A,\mathcal{G}^2_A,...,\mathcal{G}^t_A,...\}$, where $\mathcal{G}^t_A=\{\mathcal{V}^t_A,\mathcal{E}^t_A\}$ is the graph at snapshot $t$,
$\mathcal{V}^t_A$ is the node set that indicates items, and $\mathcal{E}^t_A$ is the edge set that shows the transitions between items.
%
When the user of the sequence $S_A$ acts on item $v_i$ at time $t$, an edge $e$ is established from $v_{i-1}$ to $v_i$ in the graph $\mathcal{G}^t_A$.
The local dynamic graph $\mathcal{G}^t_B$ in domain $B$ could be constructed similarly. 


\textit{\textbf{Step 2:} local dynamic graph representation.}
Then we describe how to achieve message propagations and generate the sequence embeddings in local dynamic graphs.
A common way to model dynamic graphs is to have a separate GNN handle each snapshot of the graph and feed the output of each GNN to a peephole LSTM, like GCRN \cite{seo2018structured} and DyGGNN\cite{taheri2019learning}.
Since we aim to provide a general dual graph modeling structure for CDSR, the choice of GNN and LSTM is not our focus.
%
In this special sequential setting, the challenge is how to encode the structural information as well as the sequential context of neighbors in the graph.
%
Thus, we adopt SRGNN \cite{wu2019session} here, a GNN that uses gated recurrent units (GRU) to unroll the recurrence of propagation for a fixed number of steps and then obtain the node embeddings after updating.
In this way, the message passing process decides what information to be preserved and discarded from both graphical and sequential perspectives.
After updating, the node embeddings can be denoted as $\textbf{H}^t_{al}=\{\textbf{h}^t_{al,1},\textbf{h}^t_{al,2},...,\textbf{h}^t_{al,i}\}$, where  $ \textbf{h}^t_{al,i} \in \mathbb{R}^D$ indicates the latent vector of the item $v_i$ in $\mathcal{G}^t_A$.
Later, we aggregate the item embeddings in the sequence with an attention mechanism which gives each item a specific weight that indicating their different influences to the current interest. 
Finally, the complete sequence representation is denoted as $\textbf{SE}^t_{al}$. 
%
We simply describe the propagation and representation generation mechanism of SRGNN in Appendix A and more details can be found in \cite{wu2019session}.
Similarly, in the local dynamic graph $\mathcal{G}^t_B$ of domain $B$, we obtain the node embeddings as $\textbf{H}^t_{bl}=\{\textbf{h}^t_{bl,1},\textbf{h}^t_{bl,2},...,\textbf{h}^t_{bl,j}\}$ and the sequence representation as $\textbf{SE}^t_{bl}$.

\nosection{Global dynamic graphs}
Besides intra-domain transitions, it is also necessary to model inter-domain transitions to leverage cross-domain knowledge.
To this end, we apply global dynamic graphs to capture diverse trends of user preferences in a shared feature space.
We also introduce this part in two steps.
For \textbf{\textit{step 1}}, the \textit{global dynamic graph construction} is similar as that of local dynamic graphs. 
The only difference is that we model all the items of two domains in a shared graph and the training target becomes the merged sequence, such as $S_M = \{B_1,A_1,B_2,...,A_i,B_j,...\}$.
%
%
As a user acts on the items from two domains, we add the directed edges accordingly, then consequently get a set of graphs modeling merged sequences as $\{\mathcal{G}^1_M,\mathcal{G}^2_M,...,\mathcal{G}^T_M\}$.
For \textbf{\textit{step 2}}, the way to generate the \textit{global dynamic graph representation} is the same as that of local dynamic graphs.
We denote the node embeddings in $\mathcal{G}^t_M$ as  $\textbf{H}^t_{g} = \{\textbf{h}^t_{g,1},\textbf{h}^t_{g,2},...,\textbf{h}^t_{g,k}\}$.
Then item embeddings from domain $A$ and domain $B$ in the global graph $\mathcal{G}^t_M$ could be extracted from $\textbf{H}^t_{g}$, defined as $\textbf{H}^t_{ag} = \{\textbf{h}^t_{ag,1},\textbf{h}^t_{ag,2},...,\textbf{h}^t_{ag,i}\}$ and $\textbf{H}^t_{bg} = \{\textbf{h}^t_{bg,1},\textbf{h}^t_{bg,2},...,\textbf{h}^t_{bg,j}\}$, respectively.
Finally, the sequence embeddings of the global dynamic graph from two domains can be obtained as $\textbf{SE}^t_{ag}$ and $\textbf{SE}^t_{bg}$. 

\subsubsection{Fuse attentive gate}

We then introduce how to build the transferring bridge between local graphs and global graphs.
%
%
The purposes of transferring lay in two folds : (1) users' general evolving interest patterns in local graphs and global graphs should be incorporated both in item-level and sequence-level.
(2) Due to the structured encoding technique used in graph modeling, the influence of neighbours in the graph should also be considered to avoid the loss of context information.

%
%

\begin{figure}[t]
 \begin{center}
 \includegraphics[width=\columnwidth]{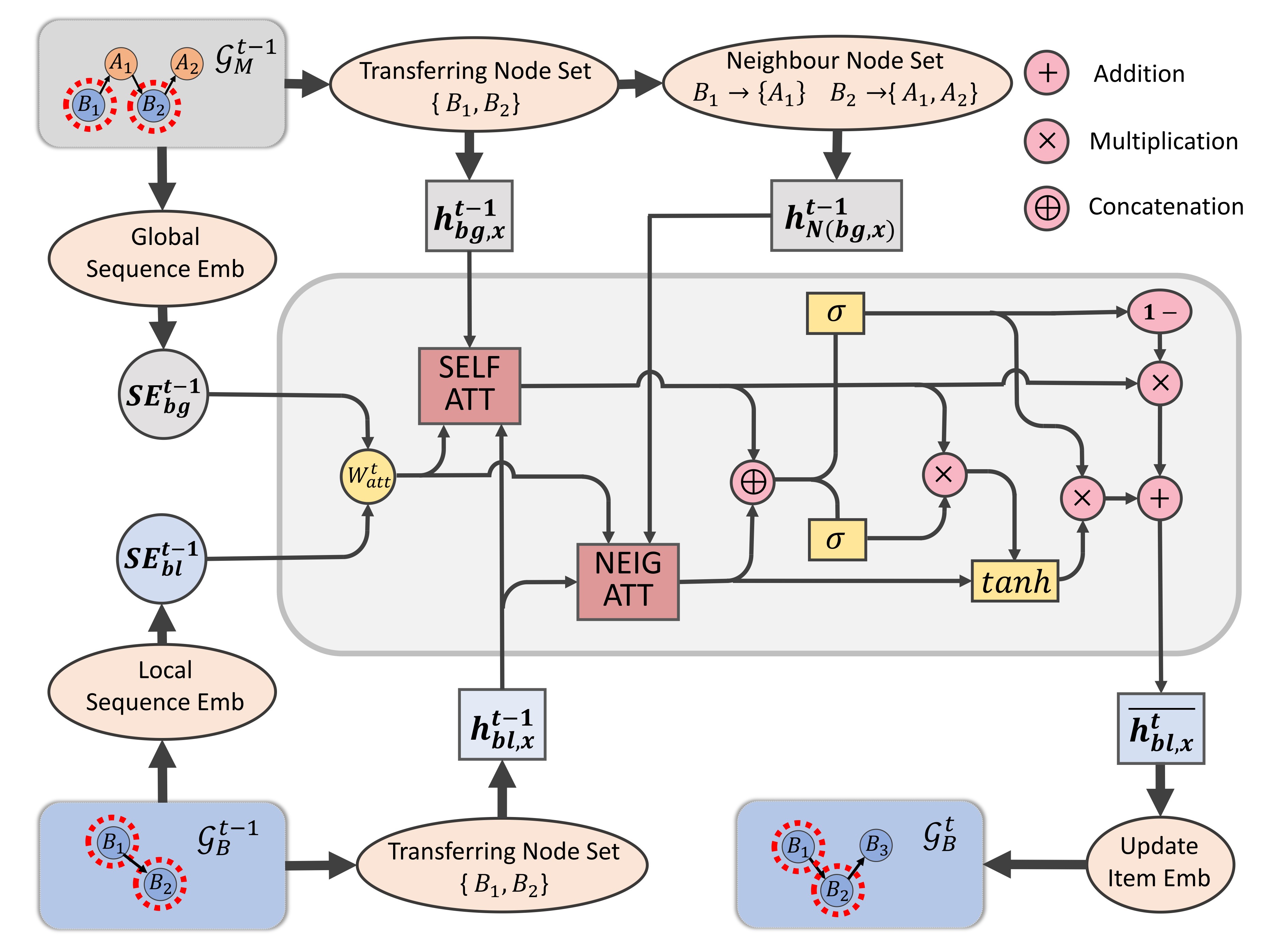}
 \caption{The structure of the fuse attentive gate. Here, we take the transfer between the global graph $M$ and the local graph of domain $B$ as an example. }
\vspace{-0.15 in}
 \label{fig:gate}
 \end{center}
\end{figure}

To achieve these purposes, we innovatively design a fuse attentive gate (FAG), which is illustrated in Figure \ref{fig:gate}.
The transferring procedure can be divided into four steps: 1) sequence-aware fusion, 2) self-attentive aggregation, 3) neighbour-attentive aggregation, and 4) updated state generation.
Here we take the merged sequence $\{B_1,A_1,B_2,A_2,B_3,...\}$ for example.
At timestamp $t$, a direct edge from $B_2$ to $B_3$ is added into the local graph $\mathcal{G}^t_B$, and we can get the transferring node set, i.e., the nodes that exist both in  $\mathcal{G}^{t-1}_B$ and $\mathcal{G}^{t-1}_M$, as $\{B_1,B_2\}$.
First, we generate the sequence embeddings on the global graph and the local graph to achieve the \textit{sequence-aware fusion}.
Second, we apply \textit{self-attentive aggregation} between $\{B_1,B_2\}$ themselves in $\mathcal{G}^{t-1}_M$ and $\mathcal{G}^{t-1}_B$.
Third, we apply neighbour-attentive aggregation between $\{B_1,B_2\}$ in $\mathcal{G}^{t-1}_B$ and their neighbour node set in $\mathcal{G}^{t-1}_M$, such as $B_1$ with its neighbour $\{A_1\}$ and $B_2$ with its neighbours $\{A_1,A_2\}$.
Finally, the results of two aggregations are integrated by a GRU unit to complete the \textit{updated state generation}.
Taking the transfer between the global graph $M$ and the local graph of domain $B$ for example, we show details of each step as follows.

\nosection{Sequence-aware fusion}
Firstly, we take the sequence embedding $\textbf{SE}^{t-1}_{bg}$ and $\textbf{SE}^{t-1}_{bl}$ generated from the global graph $\mathcal{G}^{t-1}_M$ and the local graph $\mathcal{G}^{t-1}_B$ respectively to obtain the sequence-aware attention weight: $\mathbf{W}^t_{att} = \mathbf{W}^t_{gen}[\textbf{SE}^{t-1}_{bg},\textbf{SE}^{t-1}_{bl}]+\mathbf{b}^t_{gen}$.
%
%

To efficiently and accurately fuse the structural information, we consider the feature aggregation in two perspectives, i.e., \textit{self-attentive aggregation} and \textit{neighbour-attentive aggregation}.

\nosection{Self-attentive aggregation}
The self-attentive aggregation is designed to retain the self-consistency information by transferring self-feature of all nodes in $\mathcal{G}^{t-1}_B$ from the global space to the local space.
We take the local embedding $\textbf{h}^{t-1}_{bl,x}$ of a node $x$ from $\mathcal{G}^{t-1}_B$ and its global embedding $\textbf{h}^{t-1}_{bg,x}$ from $\mathcal{G}^{t-1}_M$, then apply the self-attention:
%
\begin{equation}
\begin{aligned}
    \mathbf{\alpha}_{self-l} &= v^\top \sigma(\mathbf{W}^t_{att}\textbf{h}^{t-1}_{bl,x}+\mathbf{W}^t_{self}\textbf{h}^{t-1}_{bl,x})\\
    \mathbf{\alpha}_{self-g} &= v^\top \sigma(\mathbf{W}^t_{att}\textbf{h}^{t-1}_{bg,x}+\mathbf{W}^t_{self}\textbf{h}^{t-1}_{bl,x})\\
    \widetilde{\textbf{h}^t_{bl,x}} &= \mathbf{\alpha}_{self-l}\cdot \textbf{h}^{t-1}_{bl,x}+\mathbf{\alpha}_{self-g}\cdot \textbf{h}^{t-1}_{bg,x},
\end{aligned}
\end{equation}
where $\mathbf{\alpha}_{self-l}$ and $\mathbf{\alpha}_{self-g}$ are attention coefficients for local features and global features respectively,
and $\widetilde{\textbf{h}^t_{bl,x}}$ denotes the aggregated self-aware feature of the node $x$.

\nosection{Neighbour-attentive aggregation}
The neighbour-attentive aggregation aims to transfer context information of neighbours from the global graph to the local graph.
To do this, we generate the first-order neighbour set $N(bg,x)$ for each node $x$ in the transferring node set.
After that, we apply the neighbour-attention between the node $x$ and its first-order neighbours in the global graph :
\begin{equation}
\begin{aligned}
    \mathbf{\alpha}^i_{neig} &= \frac{exp((\textbf{h}^{t-1}_{bg,i})^\top\mathbf{W}^t_{nei}\textbf{h}^{t-1}_{bl,x})}{\sum_{i \in N(bg,x)}exp((\textbf{h}^{t-1}_{bg,i})^\top\mathbf{W}^t_{nei}\textbf{h}^{t-1}_{bl,x})}, \\
    \widehat{\textbf{h}^t_{bl,x}} &= \sum_{i \in N(bg,x)} \mathbf{\alpha}^i_{neig}\cdot\mathbf{W}^t_{att} \textbf{h}^{t-1}_{bg,i}.
\end{aligned}
\end{equation}
%
%
Here, $\mathbf{\alpha}^i_{neig}$ indicates the attention coefficients for different neighbours and $\widehat{\textbf{h}^t_{bl,x}}$ denotes the aggregated neighbour-aware feature of the node $x$.

\nosection{Updated state generation}
With $\widetilde{\textbf{h}^t_{bl,x}}$ and $\widehat{\textbf{h}^t_{bl,x}}$, we then employ a GRU-based gating mechanism as the recurrent unit:
\begin{equation}
\begin{aligned}
    z_{gate} &= \sigma(\mathbf{W}_{gz}[\widehat{\textbf{h}^t_{bl,x}},\widetilde{\textbf{h}^t_{bl,x}}]),\\
    r_{gate} &= \sigma(\mathbf{W}_{gr}[\widehat{\textbf{h}^t_{bl,x}},\widetilde{\textbf{h}^t_{bl,x}}]),\\
    \textbf{h}_{xg} &= \tanh(\mathbf{W}_{gh}[\widehat{\textbf{h}^t_{bl,x}},r_{gate}\odot \widetilde{\textbf{h}^t_{bl,x}}],\\
    \overline{\textbf{h}^t_{bl,x}} &= (1-z_{gate})\odot\widetilde{\textbf{h}^t_{bl,x}}+z_{gate}\odot \textbf{h}_{xg},
\end{aligned}
\end{equation}
where $\mathbf{W}_{gz}$, $\mathbf{W}_{gr}$, $\mathbf{W}_{gh}$ are weight matrices, and $\overline{\textbf{h}^t_{bl,x}}$ is the output of the gate.

With dual dynamic graphs and fuse attentive gates, we complete extracting item and sequence embeddings of both domains. 
The thorough algorithm of this module is show in Appendix B.

\subsection{Hybrid Metric Training}

With the final sequence representations from domain $A$ and $B$, the hybrid metric training module makes the prediction.
This module mainly has two purposes, including (1) exploring user preferences for different items in two domains, which is the prime purpose of the model, (2) alleviating the remaining data sparsity problem. 
%
For the first purpose, we apply the basic prediction loss, similar as existing researches.
To achieve the second purpose, we optimize the representation learning process with two enhancements, i.e., \textit{collaborative metric} for retaining representation alignment and \textit{contrastive metric} for preserving uniformity, to further improve the recommendation performance.

\nosection{Basic prediction}
Motivated by previous studies \cite{ma2019pi,guo2021gcn,sun2021parallel}, we calculate the matching between sequence-level embeddings $\textbf{SE}_{al,S_A}$, $\textbf{SE}_{bl,S_B}$ and item embedding matrices $\textbf{H}_{al}$, $\textbf{H}_{bl}$ in corresponding domains, to compute the recommendation probabilities:
\begin{equation}
\begin{aligned}
    P(A_{i+1}|S_A,S_B,S_M) &= softmax(\textbf{H}_{al}\cdot \textbf{SE}_{al,S_A}^\top + \textbf{b}_A),\\
    P(B_{j+1}|S_B,S_A,S_M) &= softmax(\textbf{H}_{bl}\cdot \textbf{SE}_{bl,S_B}^\top + \textbf{b}_B),
\end{aligned}
\end{equation}
%
%
%
where $\textbf{b}_A$ and $\textbf{b}_B$ are bias items.
Then the negative log-likelihood loss function is employed as follows:
\begin{equation}
\begin{aligned}
    \mathcal{L}_A &= -\frac{1}{|\mathbb{S}|}\sum_{S_A \in\mathbb{S}}\sum_{A_i \in S_A}\log P(A_{i+1}|S_A,S_B,S_M),\\
    \mathcal{L}_B &= -\frac{1}{|\mathbb{S}|}\sum_{S_B \in\mathbb{S}}\sum_{B_j \in S_B}\log P(B_{j+1}|S_B,S_A,S_M),
\end{aligned}
\end{equation}
where $\mathbb{S}$ denotes the training sequences in both domains.
%

%
%

\nosection{Collaborative metric} 
The collaborative metric aims to achieve representation alignment, which means that similar instances are expected to have similar representations.
Here, we model the observed behavior sequences as a set of positive user-item pairs $\mathcal{T}$ and learn a user-item joint metric to encode these relationships \cite{hsieh2017collaborative}. 
The learned metric tries to pull the pairs in $\mathcal{T}$ closer and push the other pairs further apart.
Specifically, we learn the model parameters by optimizing the large margin nearest neighbor objective \cite{park2018collaborative}:
\begin{equation}
\small
\begin{aligned}
    \mathcal{L}_{col} &= \sum_{(i,j)\in\mathcal{T}}\sum_{(i,k)\notin\mathcal{T}}\mathbf{W}^{rank}_{ij}\cdot max(m+D(i,j)^2-D(i,k)^2,0), 
\end{aligned}
\end{equation}
where $j$ is an item in the historical sequences of user $i$, $k$ is an item that never appears in these sequences.
$D(i,j)$ is defined as the Euclidean distance between the sequence embedding of user $i$ and the embedding of item $j$.
%
Specifically, $\mathbf{W}^{rank}_{ij}$ denotes 
the Weighted Approximate-Rank Pairwise (WARP) loss \cite{weston2010large}, calculated as $log(rank_D(i,j)+1)$ where $rank_D(i,j)$ indicates the rank of item $j$ in user $i$’s recommendations.
%
%
%
Following \cite{hsieh2017collaborative}, we empirically set the safety margin $m$ to 1.5 in our experiments.
%
%
%
%

\nosection{Contrastive metric}
The contrastive metric aims to realize representation uniformity, which means that the distribution of representations is preferred to preserve as much information as possible.
%
To do this, the contrastive loss minimizes the difference between the augmented and the original views of the same user historical sequence, and meanwhile maximizes the difference between the augmented sequences derived from different users.

Here we apply a random \textit{Item Mask} as the augmentation operator.
We take domain A for example.
For each user historical sequence $S_{A,u}$, we randomly mask a proportion of items and denote the masked sequence as $S^{mask}_{A,u}$.
%
For a mini-batch of $N$ users $\{u_1,u_2,...u_N\}$, we apply augmentation to each user's sequence and obtain 2$N$ sequences $\left[S_{A,u_1},S^{mask}_{A,u_1},S_{A,u_2},S^{mask}_{A,u_2},...,S_{A,u_n},S^{mask}_{A,u_n}\right]$. 
For each user $u$, we treat $(S_{A,u},S^{mask}_{A,u})$ as the positive pair and treat other $2(N-1)$ examples as negative samples.
We utilize dot product to measure the similarity between each representation, $sim(u,v)=u^\top v$.
Then the contrastive loss function is defined using the widely used softmax cross entropy loss as: 
\begin{equation}\small
\begin{aligned}
    \mathcal{L}_{con} = \sum_{\mathcal{P} \in \mathbb{P}} - \frac{1}{|\mathcal{P}|} \sum_{(x,y)\sim\mathcal{P}}log \frac{exp(sim(\textbf{SE}_x,\textbf{SE}_y))}{\sum_{x^{'}\in\{y\}\cup \mathcal{X}^-_\mathcal{P}}exp(sim(\textbf{SE}_x,\textbf{SE}_{x^{'}}))},
\end{aligned}
\end{equation}
where $\mathbb{P}$ is the set of positive pairs, $\mathcal{X}^-_\mathcal{P}$ is the sampled negative set for $\mathcal{P}$.

\nosection{Putting together}
Finally, we combine the basic prediction loss of both domains with collaborative and contrastive metrics via a multi-task training scheme: 
\begin{equation}
\begin{aligned}
\mathcal{L} = \mathcal{L}_A + \mathcal{L}_B + \lambda _{col}\mathcal{L}_{col} + \lambda_{con}\mathcal{L}_{con},
\end{aligned}
\end{equation}
where $\lambda_{col}$ and $\lambda_{con}$ are hyper-parameters to balance different types of losses.

\section{EXPERIMENTS AND ANALYSIS}

In this section, we conduct experiments to answer the following research questions: 
\textbf{RQ1}: How does our model perform compared with the state-of-the-art SR, CDR, and CDSR methods?
%
%
\textbf{RQ2}: How does each component of the dual dynamic graph modeling module contribute to the final performance?
%
%
\textbf{RQ3}: How does each component of the hybrid metric training module contribute to the final performance?
\textbf{RQ4}: How does our model perform with different sequence lengths, i.e., with different data sparsity degree?
\textbf{RQ5}: How do hyper-parameters affect our model performance?

\subsection{Experimental Setup} 

\nosection{Datasets}
We conduct experiments on the Amazon dataset 
\cite{mcauley2015image}, which consists of user interactions (e.g. userid, itemid, ratings, timestamps) from multiple domains.
Compared with other recommendation datasets, the Amazon dataset contains overlapped user interactions in different domain and is equipped with adequate sequential information, and thus is commonly used for CDSR research \cite{ma2019pi,guo2021gcn,li2021dual}. 
Specifically, we pick two pairs of complementary domains "\textbf{Movie} \& \textbf{Book}" and "\textbf{Food} \& \textbf{Kitchen}" for experiments.
Following the data preprocessing method of \cite{ma2019pi,ma2022mixed}, we first pick the users who have interactions in both domains and then filter the users and items with less than 10 interactions.
%
%
After that, we order and split the sequences from each user into several small sequences, with each sequence containing interactions within a period, i.e, three months for "Movie\&Book" dataset and two years for "Food\&Kitchen" dataset.
We also filter the sequences that contain less than five items in each domain.
%
The detailed statistics of these datasets are shown in Table \ref{tab:dataset}.

\begin{table}[t]
  \centering
\small
  \caption{Dataset statistics of Amazon.}
    \begin{tabular}{cccc}
    \hline
    
    \multicolumn{2}{c}{ \textbf{Movie \& Book} } & \multicolumn{2}{c}{ \textbf{Food \& Kitchen} }\\
    
    \cmidrule(lr){1-2} \cmidrule(lr){3-4}
    \textit{Movie\quad}\#Items  & 58,371 & \textit{Food\quad}\#Items & 47,520\\
    
    \cmidrule(lr){1-2} \cmidrule(lr){3-4}
    \textit{Book\quad}\#Items & 104,895 &  \textit{Kitchen\quad}\#Items & 53,361\\

    \cmidrule(lr){1-2} \cmidrule(lr){3-4}
    
    \#Overlapped-users & 12,746 & \#Overlapped-users & 8,575 \\
    
    \cmidrule(lr){1-2} \cmidrule(lr){3-4}
    
    \#Sequences & 158,373 & \#Sequences & 67,793 \\
    \#Training-sequences & 118,779 & \#Training-sequences & 50,845 \\
    \#Validation-sequences & 23,756 & \#Validation-sequences & 10,169 \\
    \#Test-sequences & 15,838 & \#Test-sequences & 6,779 \\
    
    \cmidrule(lr){1-2} \cmidrule(lr){3-4}
    Sequence Avg Length & 32.1 & Sequence Avg Length & 18.7 \\
    
\hline

    \end{tabular}%
  \label{tab:dataset}%
\end{table}%

\nosection{Evaluation method}
Following \cite{ma2019pi,sun2021parallel,guo2021gcn}, we use the latest interacted item in each sequence as the ground truth.
We randomly select 75\% of the sequences as the training set, 15\%
as the validation set, and the remaining 10\% as the test set.
We choose three evaluation metrics, i.e., Hit Rate (HR), Normalized Discounted Cumulative Gain (NDCG) and Mean Reciprocal Rank (MRR), where we set the cut-off of the ranked list as 5, 10, and 20.
For all the experiments, we repeat them five times and report the average results.

\nosection{Parameter settings}
For a fair comparison, we choose Adam \cite{kingma2014adam} as the optimizer, and tune the parameters of \modelname~and the baseline models to their best values.
Specifically, we study the effect of the hidden dimension $D$ by varying it in $\{16,32,64,128,256,512\}$, and the effects of hyper-parameters $\lambda_{col}$ and $\lambda_{con}$ by varying them in $\{0.1,0.3,0.5,0.7,1,3,5,10\}$.
%
%
And we set the batch size $N=100$.

\begin{table*}[t]
\renewcommand\arraystretch{0.6}
\centering
\small
\caption{Experimental results on Amazon datasets.}
\label{tab:compare}
\resizebox{\linewidth}{!}{
\begin{tabular}{ccccccccccccc}
\toprule
\multirow{4}{*}{}
& \multicolumn{6}{c}{\textbf{Movie-domain}}
& \multicolumn{6}{c}{\textbf{Book-domain}}\\
\cmidrule(lr){2-4} \cmidrule(lr){5-7} \cmidrule(lr){8-10} \cmidrule(lr){11-13}
&HR@5 &NDCG@5 &MRR@5 &HR@20 &NDCG@20 &MRR@20 &HR@5 &NDCG@5 &MRR@5 &HR@20 &NDCG@20 &MRR@20\\
\midrule
POP & .0107 & .0052 & .0044& .0165 & .0058 & .0047 & .0097 & .0033 & .0026& .0146 & .0041 &.0035 \\
BPR-MF & .0479 & .0402 & .0254& .0584 & .0430 & .0373 & .0374 & .0320 & .0245& .0546 & .0385&.0304\\
Item-KNN & .0745 & .0513 & .0428& .0758 & .0557  &.0456 & .0579 & .0442 &.0367 & .0663 & .0495 &.0420\\


\midrule 
GRU4REC & .2232 & .1965 & .1772 & .2456 & .2078 & .1830 & .2012 & .1648 & .1498 &.2145 & .1672 & .1520\\


SR-GNN & .2362 & .2073 & .1836 & .2592 & .2289 & .1950 & .2189 & .1733 & .1578 & .2254 & .1745 &.1611\\

BERT4Rec & .2425 & .2104 & .1989 & .2637 & .2372 & .2146 & .2203 & .1820 & .1655 & .2274 & .1834 & .1704\\

CL4SRec & .2547 &.2231 &.2035 &.2740 &.2418 &.2197 &.2325 &.1944 &.1701 &.2382 &.1987 &.1798\\

\midrule 
NCF-MLP++ & .1038 & .0679 &.0557 & .1283 & .0746 &.0675 & .0972 & .0550 & .0484 & .1080 & .0607 & .0535 \\ 

Conet & .1247 & .0742 & .0628 & .1354 & .0818 & .0732 & .1130 & .0611 & .0545 & .1205 & .0649 & .0569\\

DDTCDR & .1328 & .0933 & .0840 & .1597 & .1040 & .0996 & .1243 & .0868 &.0688 & .1548 & .0912 &.0735 \\

DARec & .1564 & .1179 & .1053 & .1796 & .1283 & .1164 & .1475 & .1089 & .0849 & .1735 & .1122 & .1004\\
\midrule 
DAT-MDI &.2453 & .2121 & .2043 & .2614 & .2340 & .2124 & .2205 & .1838 & .1676 & .2282 & .1865 &.1735\\

$\pi$-Net & .2665 & .2243 & .2108 & .2803 & .2565 & .2433 & .2385 & .2024 &.1815 & .2433 & .2134 &.2042\\

PSJNet & .2771 & .2294 &.2156 & .2917 & .2611 & .2478& .2430 & .2093 &.1879& .2609 & .2311 &.2110 \\

DASL & .2835 & .2314 & .2225 & .3012 & .2682 &.2510 & .2479 & .2133 &.1937& .2694 & .2382 & .2201\\

DA-GCN & .2821 & .2342 & .2276 & .2989 & .2719 & .2615 & .2392 & .2156 &.2038 & .2658 & .2430 & .2327\\
\midrule 
\modelname-L & .2674 &.2241 &.2139 &.2905 &.2627 &.2518 &.2435 &.2126 &.1898 &.2526 &.2219 &.2176\\
\modelname-G & .2320 &.1978 &.1744 & .2542 &.2175 &.1901 & .2092 &.1695 &.1510 &.2192 &.1709 &.1589\\
\modelname-GA & .2895 &.2423 & .2329 & .3068 & .2716 & .2648 &.2486 &.2234 &.2004 &.2673 &.2414 &.2321\\
\modelname & \textbf{.3083} & \textbf{.2517} & \textbf{.2431} & \textbf{.3257} & \textbf{.2892} & \textbf{.2734} & \textbf{.2615} & \textbf{.2379} &\textbf{.2198} &\textbf{.2890} &\textbf{.2592} &\textbf{.2448} \\
\toprule
\multirow{4}{*}{}
& \multicolumn{6}{c}{\textbf{Food-domain}}
& \multicolumn{6}{c}{\textbf{Kitchen-domain}}\\
\cmidrule(lr){2-4} \cmidrule(lr){5-7} \cmidrule(lr){8-10} \cmidrule(lr){11-13}
&HR@5 &NDCG@5 &MRR@5 &HR@20 &NDCG@20 &MRR@20 &HR@5 &NDCG@5 &MRR@5 &HR@20 &NDCG@20 &MRR@20\\

\midrule
POP &.0085 &.0048 & .0031 &.0123 &.0052 &.0035 &.0087 &.0053 &.0034 &.0152 &.0117 &.0074\\

BPR-MF &.0298 &.0237 &.0195 &.0386 &.0340 &.0255 &.0349 &.0297 &.0261 &.0422 &.0354 &.0285\\

Item-KNN &.0572 &.0389 &.0299 &.0685 &.0443 &.0396 &.0590 &.0453 &.0428 &.0702 &.0516 &.0463\\

\midrule 
GRU4REC &.1845 &.1533 &.1407 &.2076 &.1648 &.1520 &.1974 &.1608 &.1521 &.2149 &.1807 &.1685\\


SR-GNN &.1987 &.1735 &.1558 &.2196 &.1902 &.1693 &.2254 &.1969 &.1811 &. 2390 &.2078 &.1874\\

BERT4Rec &.2092 &.1824 &.1605 &.2247 &.2013 &.1745 &.2307 &.2035 &.1898 &.2432 &.2109 &.1926\\

CL4SRec &.2153 &.1918 &.1731 &.2350 &.2145 &.1876 &.2415 &.2163 &.1979 &.2582 &.2234 &.2065\\

\midrule 
NCF-MLP++ &.0801 &.0565 &.0433 &.1043 &.0672 &.0548 &.0947 &.0680 &.0573 &.1166 &.0889 &.0654\\ 

Conet &.0955 &.0730 &.0544 &.1138 &.0876 &.0685 &.1069 &.0854 &.0712 &.1244 &.1015 &.0830\\

DDTCDR &.1134 &.0975 &.0728 &.1259 &.1026 &.0759 &.1268 &.1024 &.0845 &.1398 &.1156 &.0985\\

DARec &.1210 &.1025 &.0899 &.1304 &.1118 &.0925 &.1331 &.1146 &.0933 &.1445 &.1280 &.1053\\
\midrule 
DAT-MDI  &.2123 &.1899 &.1685 &.2375 &.2144 &.1830 &.2348 &.2102 &.1956 &.2514 &.2188 &.2013\\

$\pi$-Net &.2348 &.2054 &.1834 &.2469 &.2286 &.1945 &.2575 &.2220 &.2135 &.2644 &.2290 &.2206\\ 

PSJNet &.2430 &.2118 &.1905 &.2552 &.2347 &.2039 &.2635 &.2389 &.2296 &.2718 &.2459 &.2375\\

DASL & .2572 &.2248 &.2145 &.2689 &.2430 &.2255 &.2743 &.2515 &.2483 &.2869 &.2608 &.2501\\

DA-GCN & .2524 &.2197 &.2089 &.2664 &.2415 &.2240 &.2785 &.2593 &.2495 &.2926 &.2682 &.2574\\

\midrule 
\modelname-L &.2360 &.2078 &.1897 &.2522 &.2304 &.2001 &.2599 &.2386 &.2285 &.2714 &.2448 &.2383\\
\modelname-G &.2187 &.1851 &.1644 &.2282 &.2199 &.1902 &.2404 &.2167 &.2038 &.2569 &.2235 &.2072\\
\modelname-GA &.2632 &.2348 &.2159 &.2677 &.2426 &.2250 &.2764 &.2582 &.2397 &.2816 &.2605 &.2444\\
\modelname &\textbf{.2745} &\textbf{.2431} &\textbf{.2216} &\textbf{.2789} &\textbf{.2578} &\textbf{.2364} &\textbf{.2840} &\textbf{.2627} &\textbf{.2571} &\textbf{.3043} &\textbf{.2736} &\textbf{.2660}\\
\bottomrule

\end{tabular}
}
\end{table*}

\subsection{Comparison Methods}
\nosection{Traditional recommendations}
\textbf{POP} is the simplest baseline that ranks items according to their popularity judged by the number of interactions.
\textbf{BPR-MF} \cite{rendle2012bpr} optimizes the matrix factorization with implicit feedback using a pairwise ranking loss.
\textbf{Item-KNN} \cite{sarwar2001item} recommends the items that are similar to the previously interacted items in the sequence, where similarity is defined as the cosine similarity between the vectors of sequences.

\nosection{Sequential recommendations}
\textbf{GRU4REC} \cite{hidasi2015session} uses GRU to encode sequential information and employs ranking-based loss function.
%
%
\textbf{BERT4Rec} \cite{sun2019bert4rec} adopts a bi-directional transformer to extract the sequential patterns.
\textbf{SRGNN} \cite{wu2019session} obtains item embeddings through a gated GNN layer and then uses a self-attention mechanism to compute the session level embeddings.
\textbf{CL4SRec} \cite{xie2020contrastive} uses item cropping, masking, and reordering as augmentations for contrastive learning on interaction sequences.

\nosection{Cross-Domain Recommendations}
\textbf{NCF-MLP++} is a deep learning based method which uses multilayer perceptron (MLP) \cite{he2017neural} to learn the inner product in the traditional collaborative filtering process.
%
%
We adopt the implementation in \cite{ma2019pi}. 
\textbf{Conet} \cite{hu2018conet} enables dual knowledge transfer across domains by introducing a cross connection unit from one base network to the other and vice versa.
\textbf{DDTCDR} \cite{li2020ddtcdr} introduces a deep dual transfer network that transfers knowledge with orthogonal transformation across domains.
\textbf{DARec} \cite{yuan2019darec} transfers knowledge between domains with shared users, learning shared user representations across different domains via domain adaptation technique. 

\nosection{Cross-domain sequential recommendations}
\textbf{$\pi$-Net} \cite{ma2019pi} proposes a novel parallel information-sharing network with a shared account filter and a cross-domain transfer mechanism to simultaneously generate sequential recommendations for two domains.
\textbf{PSJNet} \cite{sun2021parallel} is a variant of $\pi$-Net, which learns cross-domain representations by extracting role-specific information and combining useful user
behaviors. 
\textbf{DA-GCN} \cite{guo2021gcn} devises a domain-aware graph convolution network with attention mechanisms to learn user-specific node representations under the cross-domain sequential setting.
\textbf{DAT-MDI} \cite{chen2021dual} uses a potential mapping method based on a slot attention mechanism to extract the user’s sequential preferences in both domains. 
\textbf{DASL} \cite{li2021dual} is a dual attentive sequential learning model, consisting of dual embedding and dual attention modules to match the extracted embeddings with candidate items. 


\subsection{Model Comparison (for RQ1)}

\nosection{General comparison} We report the comparison results on two datasets \textbf{Movie} \& \textbf{Book} and \textbf{Food} \& \textbf{Kitchen} in Table \ref{tab:compare}.
For space save, we report the results when the cut-off of the ranked list is 10 in Appendix C.
From the results, we can find that: (1) \modelname~outperforms \textbf{SR} baselines, indicating that our model can effectively capture and transfer the cross-domain information to promote the recommendation performance of a single domain.
(2) Comparing \modelname~with \textbf{CDR} baselines, the improvement in both domains on two datasets is also evident.
It proves the effectiveness of \modelname~ in dynamically modeling the intra-domain and the inter-domain sequential transitions with a graphical framework.
(3) Compared with \textbf{CDSR} baselines, \modelname~achieves better performance in terms of all metrics, which demonstrates that \modelname~is able to extract user preferences more accurately by leveraging complementary information from both domains.

\begin{table}[t]
\renewcommand\arraystretch{0.6}
\small
\centering
\caption{Model performances of different sequence lengths.}
\label{tab:length}
\begin{tabular}{ccccccc}
\toprule
\multirow{1}{*}{}
& \multicolumn{3}{c}{\textbf{Movie-domain}}
& \multicolumn{3}{c}{\textbf{Book-domain}}\\
\cmidrule(lr){2-4}
\cmidrule(lr){5-7}
\multirow{1}{*}{}
& L=10.7 & L=21.5 & L=32.1 & L=10.7 & L=21.5 & L=32.1 \\
\midrule 
CL4SRec &.1192 & .1848& .2740& .1063& .1679& .2382\\
DARec & .0925& .1139& .1796& .0832& .1313& .1735\\
DASL & .1352  & \underline{.2528} & \underline{.3012} & .1372 & \underline{.2128} & \underline{.2694}\\
DA-GCN & \underline{.1467}	& .2496 & .2989 & \underline{.1391}  & .2119 & .2658\\
\midrule
\modelname & \textbf{.1635}	& \textbf{.2766} & \textbf{.3257} & \textbf{.1548} & \textbf{.2342} & \textbf{.2890}\\
\midrule
Imp. & 11.5\% & 9.41\% & 8.13\% &11.3\% & 10.1\%& 7.28\%\\
\bottomrule
\end{tabular}
\vspace{-0.1cm} 
\end{table}

\subsection{In-depth Analysis (for RQ2-RQ5)}

\nosection{Study of the dual dynamic graph modeling (RQ2)}
To study how each component of the dual dynamic graph modeling module contributes to the final performance, we compare \modelname~with its several variants, including \modelname-L, \modelname-G, and \modelname-GA.
(a) \modelname-L applies only the local graphs in dual dynamic graph modeling to extract item and sequence embeddings. 
(b) \modelname-G applies only the global graph.
(c) \modelname-GA retains the dual dynamic graph structure, but replaces FAG with a simple GRU as cross-domain transfer unit (CTU) \cite{ma2019pi}.
The comparison results are also shown in Table \ref{tab:compare}. 
From it, we can observe that: (1) Both \modelname-L and \modelname-G perform worse than \modelname, showing that integrating intra-domain transitions with inter-domain 
transitions can boost the performance. 
(2) \modelname-L outperforms \modelname-G on two datasets, indicating that intra-domain collaborative influences still take the dominant place when extracting user preferences.
And this is because that simply encoding inter-domain transitions in a shared latent space ignores the domain feature distribution bias.
(3) \modelname~shows superiority over \modelname-GA, which proves that the FAG completes transferring more effectively than other CTUs in existing research and is especially applicable in the dual dynamic graphical model.

\nosection{Study of the hybrid metric training (RQ3)}
In order to verify the effectiveness of each component of the hybrid metric training module, we further conduct more ablation studies, where (a) \modelname-col only employs collaborative metric, (b) \modelname-con only applies contrastive metric, and (c) \modelname-cc trains the recommendation model without either metric.
The results are shown in Table \ref{tab:constraint}, from which we can conclude that: (1) Both \modelname-con and \modelname-col perform worse than \modelname~. 
The reasons are two-folds. On the one hand, 
employing collaborative metric loss augments the similarity of similar instance embeddings thus achieving representation alignment. 
On the other hand, adopting contrastive metric loss preserves the discrepancy of different feature distributions thus retaining representation uniformity.
Both of them play an important role in enhancing the representations.
(2) \modelname~significantly outperforms \modelname-cc,  which indicates that the combination of two metrics effectively optimizes the training module and further promotes model performance.

\begin{table}[t]
\renewcommand\arraystretch{0.6}
\small
\centering
\caption{Ablation results on two Amazon datasets.}
\label{tab:constraint}
\begin{tabular}{ccccccc}
\toprule
\multirow{1}{*}{}
& \multicolumn{2}{c}{\textbf{Movie-domain}}
& \multicolumn{2}{c}{\textbf{Book-domain}}\\
\cmidrule(lr){2-3}
\cmidrule(lr){4-5}
\multirow{1}{*}{}
& HR@20 & NDCG@20 & HR@20 & NDCG@20 \\
\midrule 
\modelname-col &.3159 &.2812 &.2786 &.2521 \\
\modelname-con &.3205 &.2848 &.2835 &.2554\\
\modelname-cc &.3045 &.2786 &.2704 &.2489\\
\modelname & \textbf{.3257} & \textbf{.2892} & \textbf{.2890} & \textbf{.2592}\\
\midrule
\multirow{1}{*}{}
& \multicolumn{2}{c}{\textbf{Food-domain}}
& \multicolumn{2}{c}{\textbf{Kitchen-domain}}\\
\cmidrule(lr){2-3}
\cmidrule(lr){4-5}
\multirow{1}{*}{}
& HR@20 & NDCG@20  & HR@20 & NDCG@20 \\
\midrule 
\modelname-col &.2740 &.2518 &.2959 &.2698 \\
\modelname-con &.2764 &.2540 &.2998 &.2704 \\
\modelname-cc &.2712 &.2473 &.2958 &.2695\\
\modelname &\textbf{.2789} &\textbf{.2578} &\textbf{.3043 }&\textbf{.2736}\\
\bottomrule
\end{tabular}
\vspace{-0.2cm} 
\end{table}

\begin{figure} 
    \centering
        \subfigure[\textbf{Movie} \& \textbf{Book}]{
    \begin{minipage}[t]{0.47\linewidth} 
    \includegraphics[width=4.2cm]{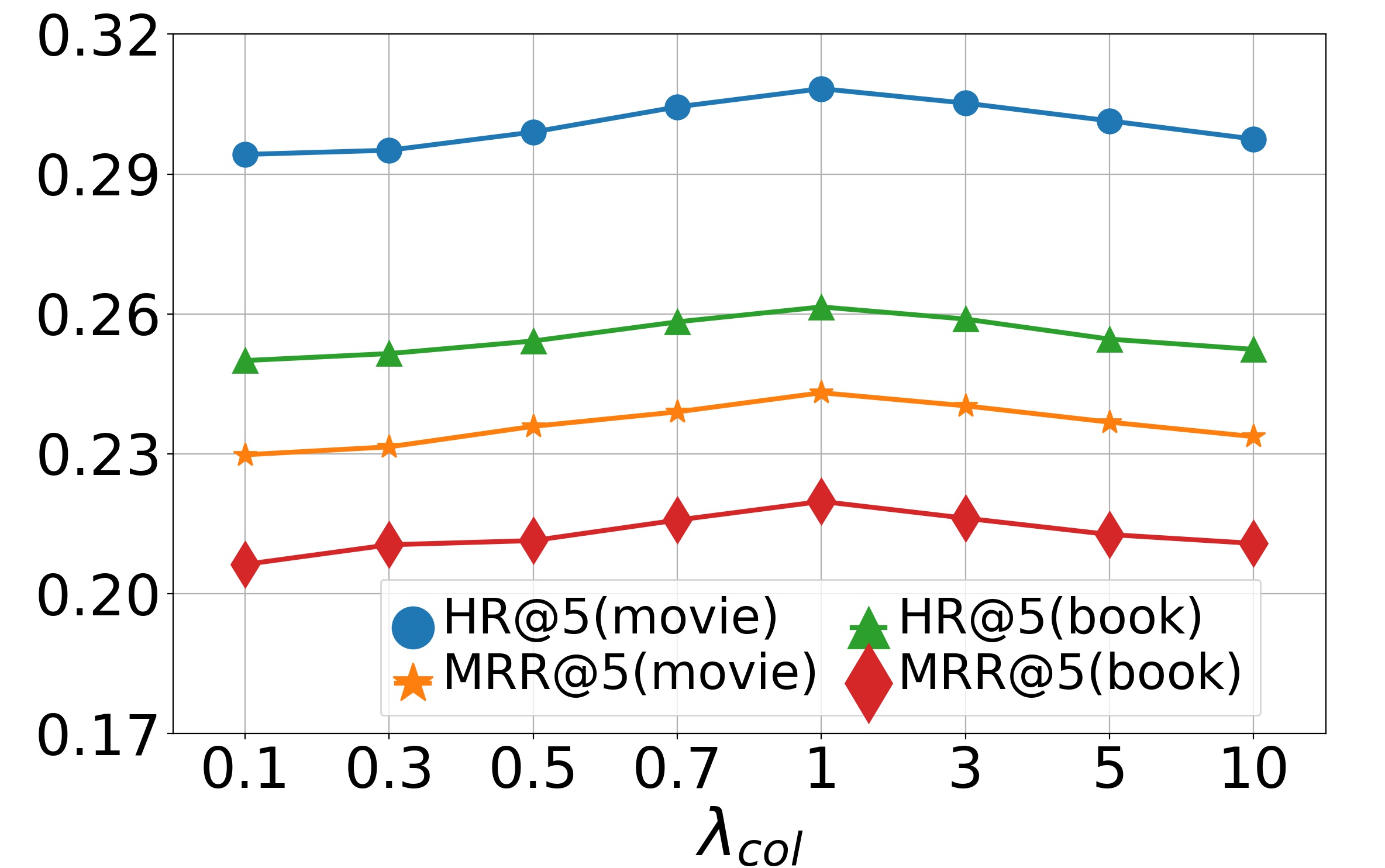}
    \end{minipage}
}
        \subfigure[\textbf{Food} \& \textbf{Kitchen}]{
    \begin{minipage}[t]{0.47\linewidth} 
    \includegraphics[width=4.2cm]{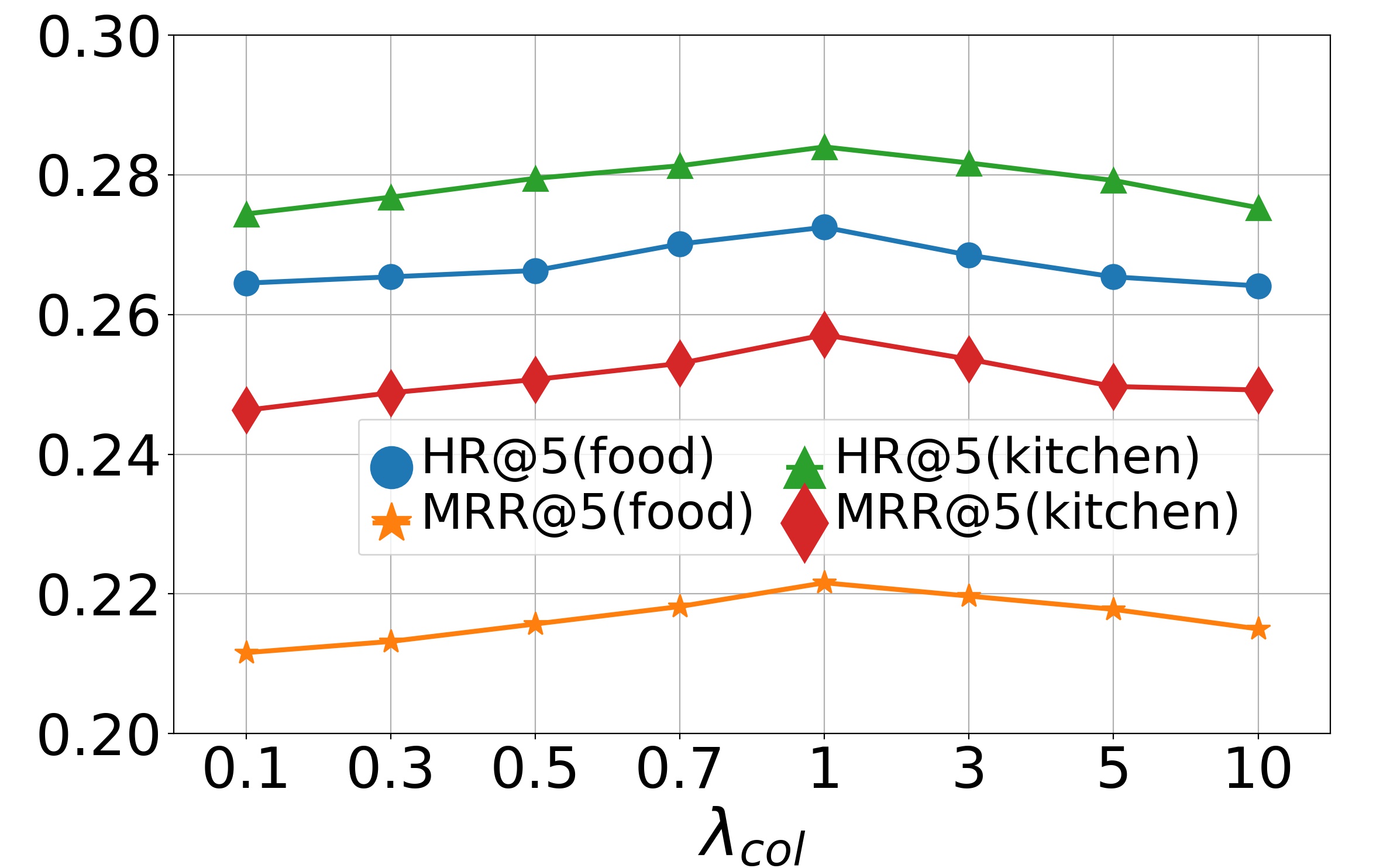}
    \end{minipage}
}
\vspace{-0.15cm}
	  \caption{Effect of $\lambda_{col}$ on \modelname.}
	  \label{fig1:col}
\vspace{-0.4cm}
\end{figure}

\begin{figure} 
    \centering
        \subfigure[\textbf{Movie} \& \textbf{Book}]{
    \begin{minipage}[t]{0.47\linewidth} 
    \includegraphics[width=4.2cm]{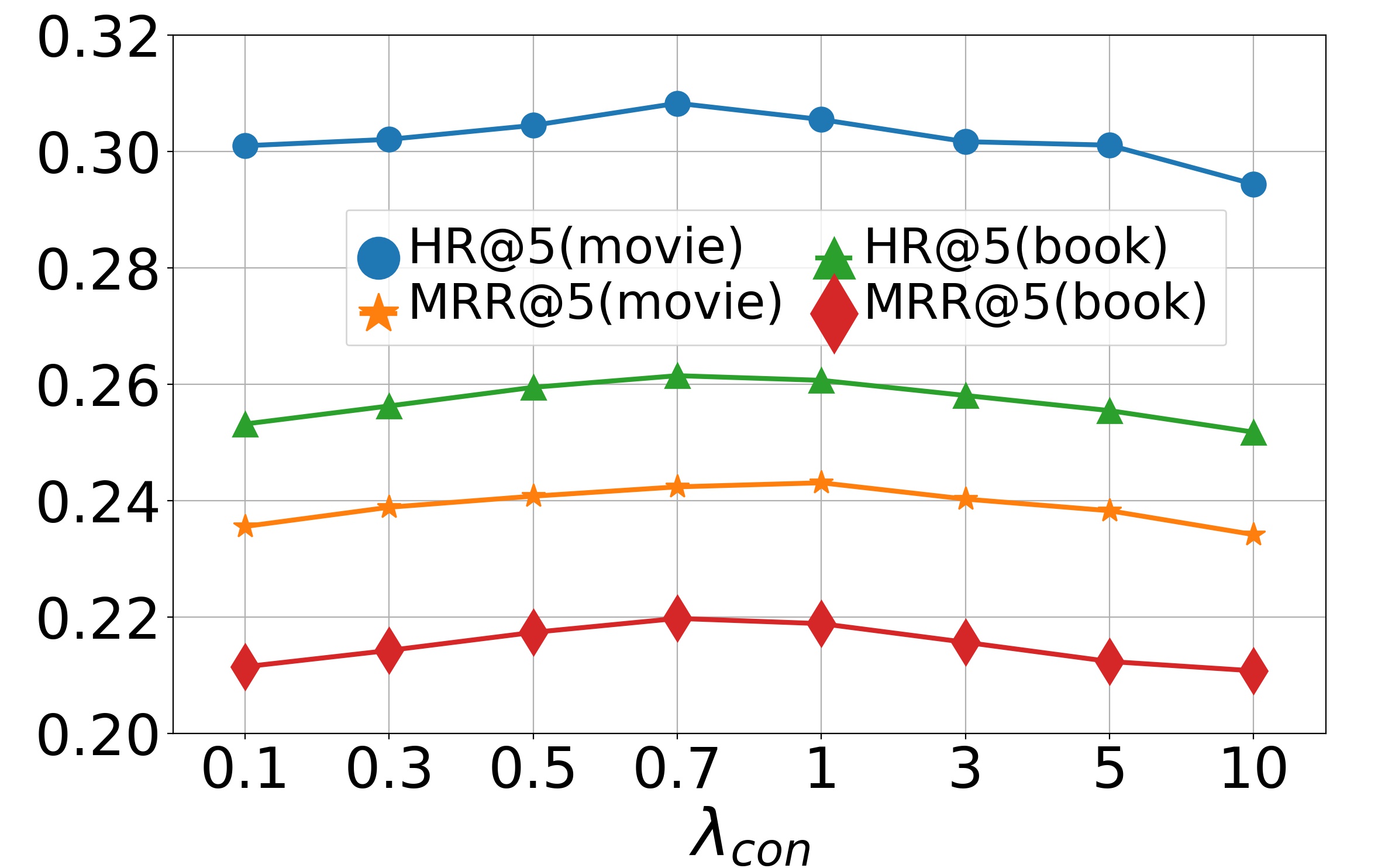}
    \end{minipage}
}
        \subfigure[\textbf{Food} \& \textbf{Kitchen}]{
    \begin{minipage}[t]{0.47\linewidth} 
    \includegraphics[width=4.2cm]{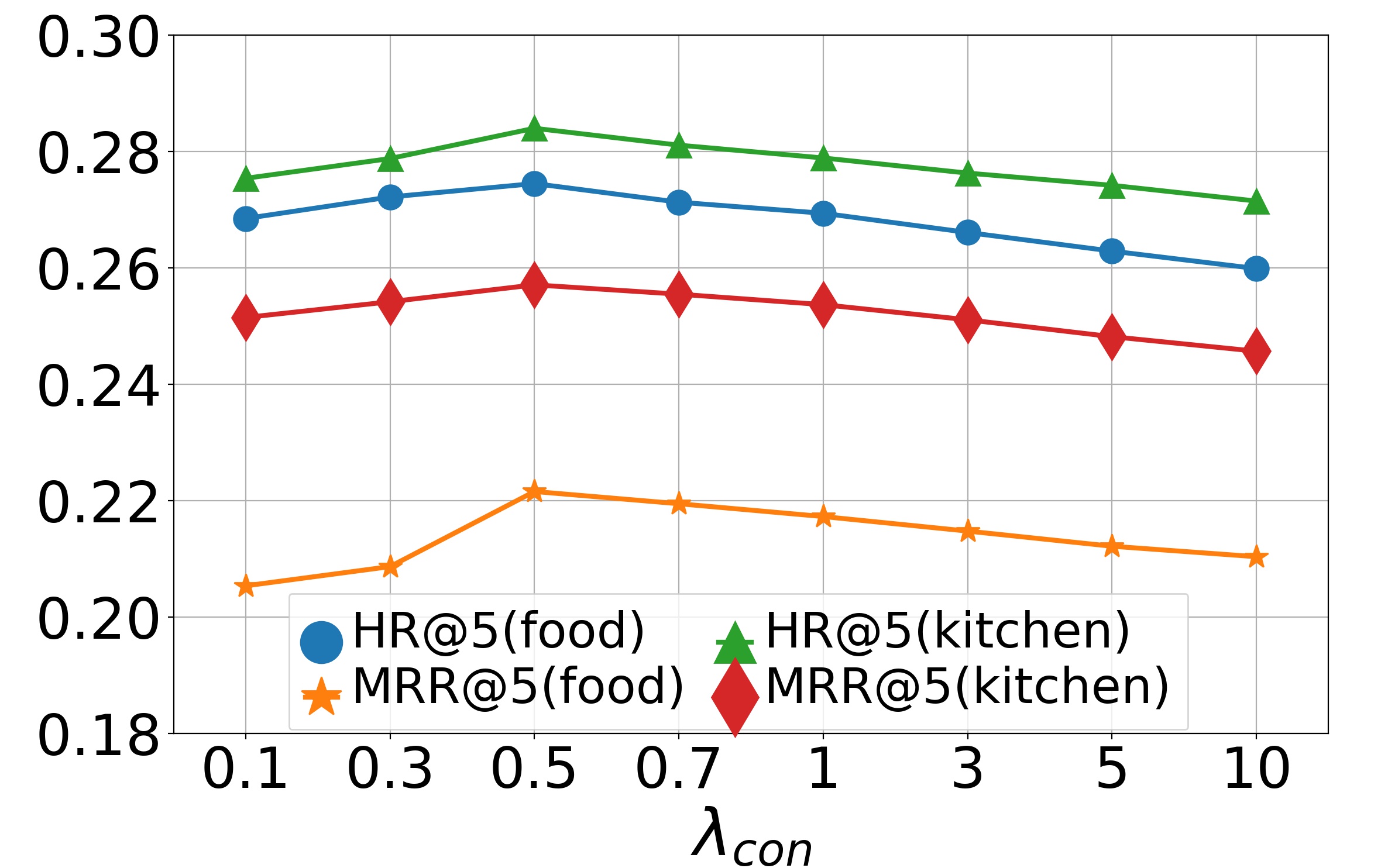}
    \end{minipage}
}
\vspace{-0.15cm}
	  \caption{Effect of $\lambda_{con}$ on \modelname.}
	  \label{fig2:con}
\vspace{-0.3cm}
\end{figure}

\begin{figure} 
    \centering
        \subfigure[\textbf{Movie} \& \textbf{Book}]{
    \begin{minipage}[t]{0.47\linewidth} 
    \includegraphics[width=4.2cm]{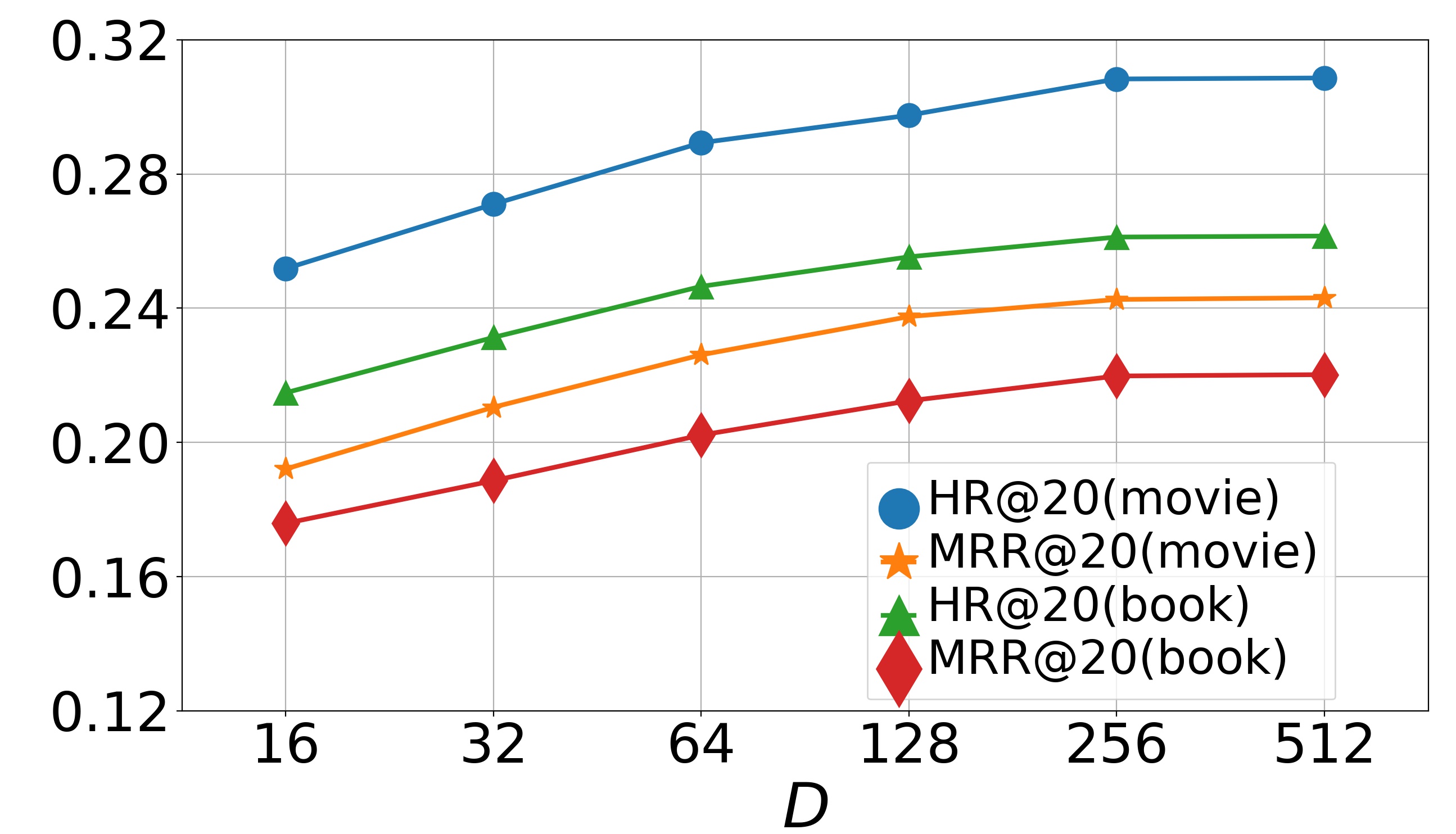}
    \end{minipage}
}
        \subfigure[\textbf{Food} \& \textbf{Kitchen}]{
    \begin{minipage}[t]{0.47\linewidth} 
    \includegraphics[width=4.2cm]{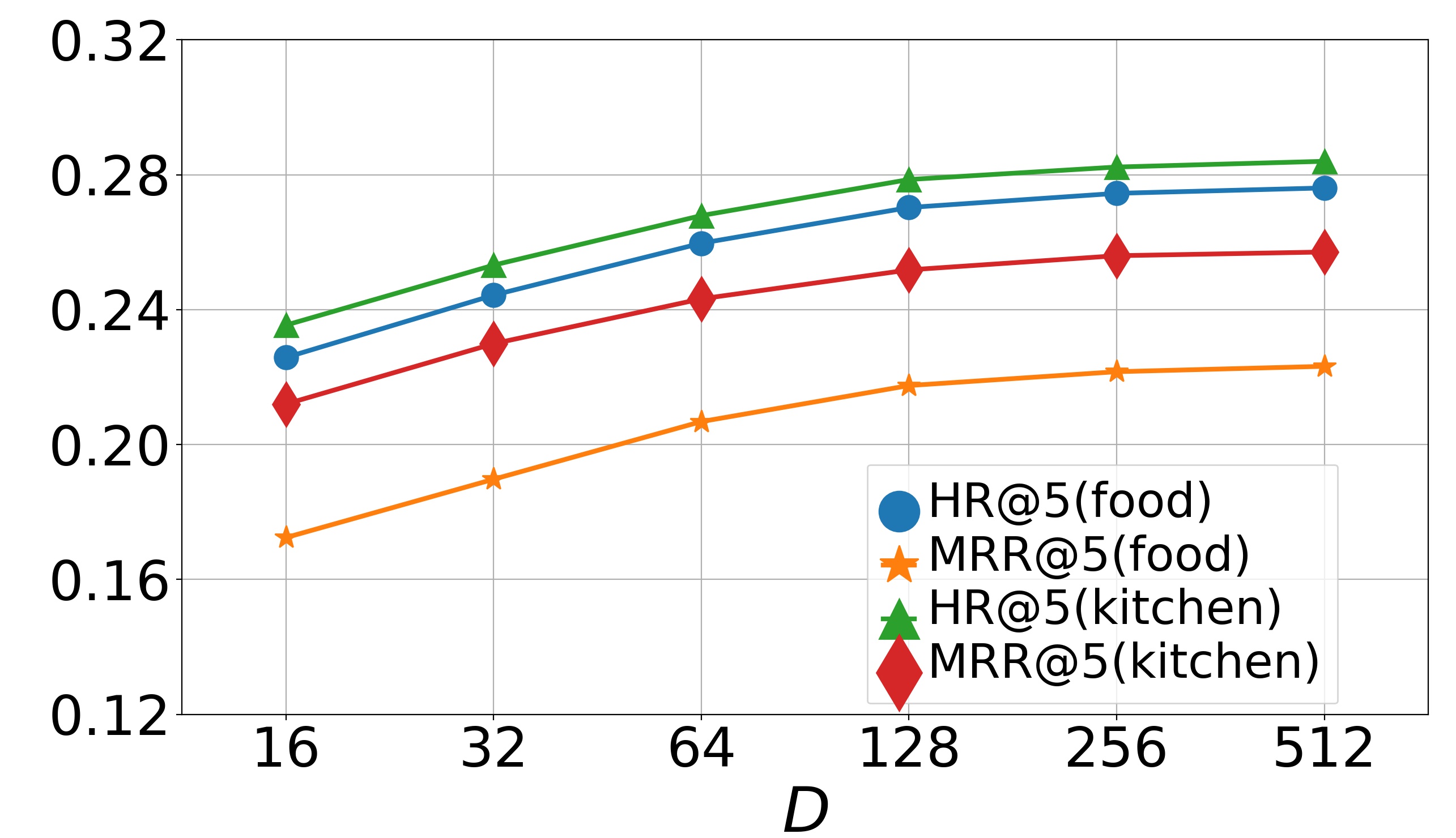}
    \end{minipage}
}
\vspace{-0.15cm}
	  \caption{Effect of embedding dimension $D$ on \modelname.}
	  \label{fig3:n}
\vspace{-0.4cm}
\end{figure}

\nosection{Influence of sequence length (RQ4)}
We report the comparison results of HR@20 under different sequence average length settings in Table \ref{tab:length}.
Here, for space save, we only report the results of several baselines which perform best in each category, i.e., SR, CDR, and CDSR.
And $L$ means the average length of sequences.
From it, we can observe that in both domains, the shorter the average length of sequences is, the grater improvement our model brings.
Therefore, \modelname~is able to effectively alleviate the data sparsity problem and performs well even when sequences are short.

\nosection{Parameter analysis (RQ5)}
We now study the effects of hyper-parameters on model performance, including $\lambda_{col}$, $\lambda_{con}$, and $D$.
We first study the effects of $\lambda_{col}$ and $\lambda_{con}$ on the model, varying them in $\{0.1,0.3,0.5,0.7,1,3,5,10\}$, and then report the results in Fig.~\ref{fig1:col} and Fig.~\ref{fig2:con}.
The bell-shaped curves show that the accuracy will first gradually increase with $\lambda_{col}$ or $\lambda_{con}$ and then slightly decrease.
We can conclude that when $\lambda_{col}$ and $\lambda_{con}$ approach 0, the collaborative loss and contrastive loss cannot produce positive effects.
But when $\lambda_{col}$ and $\lambda_{con}$ become too large, the metric loss will suppress  the negative log-likelihood loss, which also reduces the recommendation accuracy. 
Empirically, we choose $\lambda_{col}=1.0$ and $\lambda_{con}=0.7$ on the dataset \textbf{Movie} \& \textbf{Book} while $\lambda_{col}=1.0$ and $\lambda_{con}=0.5$ on the dataset \textbf{Food} \& \textbf{Kitchen}.
Finally we perform experiments by varying dimension $D$ in range $\{16,32,64,128,256,512\}$.
The results on two datasets are shown in Fig.~\ref{fig3:n}.
From them, we can see that the performance gradually improves when $D$ increases and finally keeps a fairly stable level after $D$ reaches 256.
It indicates that a larger embedding dimension can provide more accurate embeddings for items thus enriching the representations of user preferences. 
Since a too large embedding dimension will cause huge computational and time cost, we choose $D=256$ here for both datasets.

\section{Conclusion}
In this paper, we propose \modelname~, which includes a dual dynamic graph modeling module and a hybrid metric training module, for solving the Cross-Domain Sequential Recommendation (CDSR) problem.
In the dual dynamic graph modeling module, we firstly construct dual dynamic graphs, i.e., global graphs and local graphs, to explore intra-domain and inter-domain transitions, and then adopt a fuse attentive gating mechanism to adaptively integrate them.
In the hybrid metric training module, we apply the representation enhancement from two aspects, i.e., collaborative metric for alignment and contrastive metric for uniformity, so that the remaining data sparsity impact in CDSR is alleviated. 
Extensive experiments conducted on two real-world datasets illustrate the effectiveness of \modelname~and the contribution of each component.

\begin{acks}
This work was supported in part by the National Natural Science Foundation of China (No.62172362 and No.72192823) and Leading Expert of ``Ten Thousands Talent Program'' of Zhejiang Province (No.2021R52001).
\end{acks}

\bibliographystyle{ACM-Reference-Format}
\bibliography{sample-base}

\clearpage

\appendix

\section{propagation mechanism of SRGNN}\label{sec:prop}

\begin{figure}[t]
 \begin{center}
 \includegraphics[width=\columnwidth]{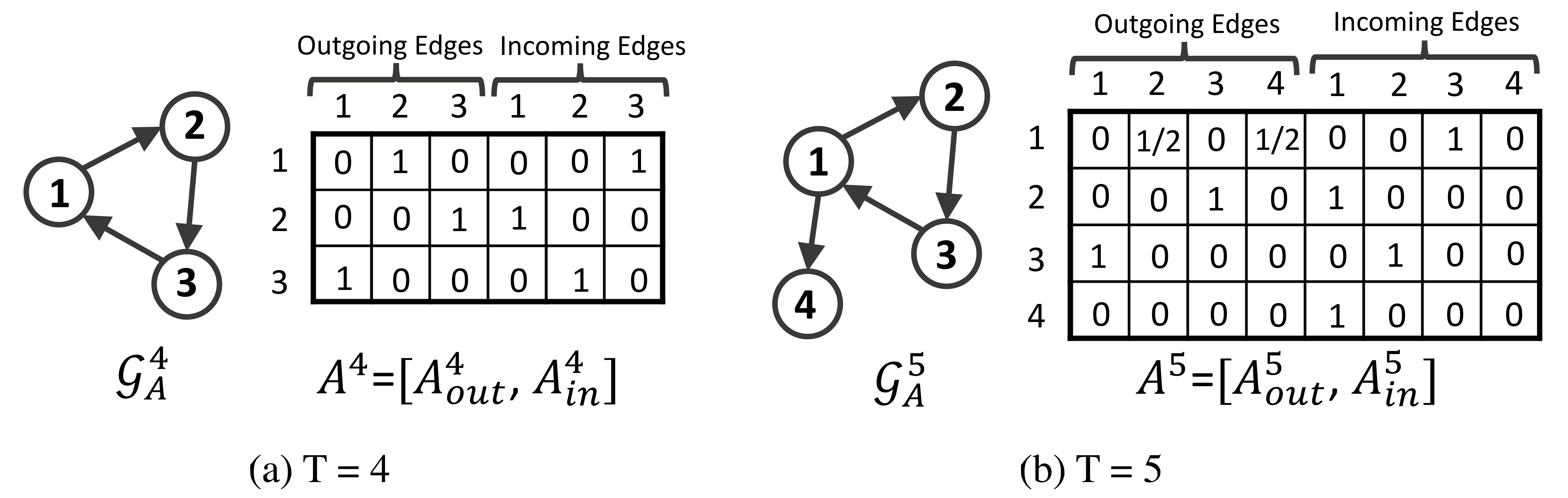}
 \caption{(a) and (b) are dynamic graphs and corresponding connection matrices.}
 \label{fig:example-a}
 \end{center}
\end{figure}

In the local dynamic graphs and the global dynamic graph, we adopt SRGNN \cite{wu2019session} to achieve message propagation.
Here, we take the local dynamic graph of the domain $A$ for example, a sequence in it can be represented by $S_A = [v_1,v_2,...,v_i,...]$, where $v_i$ represents a consumed item of the user within the sequence $S_A$.
We embed every item $v$ in domain $A$ into a unified embedding space and use a node vector $ \textbf{h}_{al} \in \mathbb{R}^D$ to denote the latent vector of the item $v$ learned via local dynamic graphs in domain $A$, with $D$ denoting the dimensionality.
For the node $v_{i}$ of the graph $\mathcal{G}^t_A$, the recurrence of the propagation functions is given as follows:
\begin{align}
\label{equ:prop}
a^k_{i} &= \textbf{A}^t_{i:}[\textbf{h}^{k-1}_{al,1},...,\textbf{h}^{k-1}_{al,n}]^\top\mathbf{W}_h+\mathbf{b}_h,\\
\label{equ:zgate}
z^k_{i} &= \sigma(\mathbf{W}_z a^k_{i}+\mathbf{U}_z \textbf{h}^{k-1}_{al,i}),\\
\label{equ:rgate}
r^k_{i} &= \sigma(\mathbf{W}_r a^k_{i}+\mathbf{U}_r \textbf{h}^{k-1}_{al,i}),\\
\label{equ:h4}
\widetilde{\textbf{h}^k_i} &= \tanh(\mathbf{W}_o a^k_{i}+\mathbf{U}_o(r^k_{i} \odot \textbf{h}^{k-1}_{al,i})),\\
\label{equ:h5}
\textbf{h}^k_{al,i} &= (1-z^k_{i})\odot \textbf{h}^{k-1}_{al,i}+z^k_{i}\odot\widetilde{\textbf{h}^k_i},
\end{align}
where $[\textbf{h}^{k-1}_{al,1},...,\textbf{h}^{k-1}_{al,n}]$ is the node embedding list in the current sequence and $k$ denotes the propagation step.
The matrix $\textbf{A}^t \in \mathbb{R}^{n \times 2n}$ is the concatenation of two adjacency matrices $\textbf{A}^t_{in}$ and $\textbf{A}^t_{out}$, which represents weighted
connections of outgoing and incoming edges in the $\mathcal{G}^t_A$ at snapshot $t$, respectively.
And $\textbf{A}^t_{i:}$ are the two columns in $\textbf{A}^t$ corresponding to node $v_{i}$. which describes how nodes in the graph communicate with each other.
For example, consider a sequence $S_A = [v_1,v_2,v_3,v_1,v_4,...]$, the dynamic graphs $\mathcal{G}^t_A$ and the corresponding matrix $\textbf{A}^t$ are illustrated in Figure. \ref{fig:example-a}.
%
We select the snapshot $t=4$ and $t=5$ as examples, and every column in $\textbf{A}^t$ is composed of the normalized outgoing and incoming edge weights.



\begin{table*}[t]
\small
\centering
\caption{Experimental results on Amazon datasets when the cut-off of the ranked list is 10.}
\label{tab:compare-1}
\begin{tabular}{ccccccccccccc}
\toprule
\multirow{3}{*}{}
& \multicolumn{3}{c}{\textbf{Movie-domain}}
& \multicolumn{3}{c}{\textbf{Book-domain}}
& \multicolumn{3}{c}{\textbf{Food-domain}}
& \multicolumn{3}{c}{\textbf{Kitchen-domain}}\\
\cmidrule(lr){2-4} \cmidrule(lr){5-7} \cmidrule(lr){8-10} \cmidrule(lr){11-13}
&HR &NDCG &MRR &HR &NDCG &MRR &HR &NDCG &MRR &HR &NDCG &MRR\\
\midrule
POP & .0125 & .0056 & .0045 & .0115 & .0038 & .0031 & .0097 & .0049 &.0033 &.0128 &.0089 &.0050\\
BPR-MF & .0513 & .0418 & .0316 & .0457 & .0359 & .0266 &  .0345 &.0278 &.0214 &.0382 &.0319 & .0435\\
Item-KNN & .0752 & .0527 &.0433 & .0604 &.0468 &.0395 & .0614 &.0412 &.0341 &.0628 &.0497 &.0443\\

\midrule 
GRU4REC & .2320 & .2014 &.1785 &.2068 &.1651 &.1510 & .1910 &.1587 &.1482 &.2034 &.1745 &.1591\\


SR-GNN & .2468 &.2159 &.1923 &.2226 &.1738 &.1599 & .2048 &.1834 &.1569 &.2298 &.2014 &.1856\\

BERT4Rec & .2510 & .2248 &.2034 &.2245 &.1825 &.1673 & .2115 &.1932 &.1683 &.2385 &.2084 &.1910\\

CL4SRec &.2656 &.2350 &.2147 &.2352 &.1964 &.1740 &.2235 &.2041 &.1775 &.2487 &.2193 &.2044\\

\midrule 
NCF-MLP++ & .1142 & .0718 &.0614 &.1015 &.0578 &. 0501 & .0946 &.0622 &.0498 &.1057 &.0754 &.0602\\ 

Conet & .1287 &.0788 &.0725 &.1182 &.0624 &.0550 & .1043 &.0795 &.0592 &.1138 &.0940 &.0775\\

DDTCDR & .1433 &.0984 &.0936 &.1327 &.0893 &.0712 & .1182 &.0980 &.0736 &.1299 &.1087 &.0923\\

DARec & .1635 &.1223 &.1099 &.1608 &.1097 &.0922 & .1278 &.1074 &.0906 &.1396 &.1221 &.0981\\
\midrule 
DAT-MDI & .2526 &.2235 &.2097 &.2247 &.1848 &.1698 &.2237 &.2054 &.1728 &.2458 &.2163 &.1989\\

$\pi$-Net & .2744 &.2359 &.2218 &.2402 &.2075 &.1903 &.2405 &.2128 &.1886 &.2616 &.2247 &.2177\\

PSJNet & .2862 &.2429 &.2347 &.2516 &.2161 &.2035 &.2512 &.2233 &.1964 &.2682 &.2412 &.2326\\

DASL & .2940 &.2587 &.2412 &.2586 &.2248 &.2159 &.2620 &.2318 &.2174 &.2812 &.2540 &.2493\\

DA-GCN & .2925 &.2624 &.2477 &.2543 &.2294 &.2198 &.2589 & .2294 &.2138 &.2856 &.2617 &.2533\\
\midrule 
\modelname-L & .2786 & .2425 &.2297 &.2468 &.2183 &.2099 &.2415 &.2187 &.1934 &.2655 &.2410 &.2330\\
\modelname-G & .2495 &.2133 &.1862 &.2144 &.1701 &.1547 &.2275 &.2089 &.1749 &.2488 &.2208 &.2062\\
\modelname-GA & .2957 &.2688 &.2534 &.2590 &.2316 &.2239 & .2658 &.2378 &.2204 &.2781 &.2599 &.2423\\
\modelname & \textbf{.3148} &\textbf{.2745} &\textbf{.2630} &\textbf{.2754} &\textbf{.2418} &\textbf{.2343} & \textbf{.2762} &\textbf{.2515} &\textbf{.2284} &\textbf{.2931} &\textbf{.2689} &\textbf{.2642}\\

\bottomrule

\end{tabular}

\end{table*}

%
Eq. \eqref{equ:prop} shows the step that passes information between different nodes of the graph via edges in both directions. 
After that, a GRU-like update, including update gate $\mathbf{z}_{i}$ in Eq. \eqref{equ:zgate} and reset gate $\mathbf{r}_{i}$ in Eq. \eqref{equ:rgate}, is adopted to determine what information to be preserved and discarded respectively. 
$\sigma(\cdot)$ is the sigmoid function and $\odot$ is the element-wise multiplication operator. 
Then the candidate state is generated by the previous state and the current state under the control of the reset gate in Eq. \eqref{equ:h4}.
We combine the previous hidden state with the candidate state using updating mechanism and get the final state in Eq. \eqref{equ:h5}.

\begin{algorithm}[t]
\caption{Dual Dynamic Graph Modeling }\label{alg:dual-dynamic-graph-algo}
\KwIn{user's cross-domain behavior sequence $S_M$;single-domain behavior sequences: $S_A$,$S_B$.}
\KwOut{Item embeddings in local domain graphs : $\textbf{H}_{al}$, $\textbf{H}_{bl}$; Sequence embeddings : $\textbf{SE}_{al,S_A}$, $\textbf{SE}_{bl,S_B}$ }
{
Initialize node embeddings in local graphs as $\textbf{H}^0_{al}$ and $\textbf{H}^0_{bl}$.\\
Global node embeddings $\textbf{H}^0_g = \textbf{H}^0_{al} \oplus \textbf{H}^0_{bl}$ \\
\textbf{for} $t=1$ to $T$ \textbf{do}\\
    \quad Get the current item $v_{i}$, $v_{i}$'s domain as $X$; previous item in $S_M$ as $v^M_{i-1}$, in $S_X$ as $v^X_{i-1} $; previous item set as $V_{pre}$.\\
    \quad \textbf{From local graph $\mathcal{G}^{t-1}_X$:} \\
    \quad $Emb(V_{pre})=\textbf{h}^{t-1}_{xl,pre}$.\\
    \quad Get local sequence embedding $\textbf{SE}^{t-1}_{xl}$ with Eq. (6-7).\\
     \quad \textbf{From global graph $\mathcal{G}^{t-1}_M$:} \\
    \quad Select $V_{pre}$'s neighbor set as $N_{xg}$.\\
    \quad $Emb(V_{pre})=\textbf{h}^{t-1}_{xg,pre}$, $Emb(N_{xg})=\textbf{h}^{t-1}_{xg,neig}$.\\
    \quad Get global sequence embedding $\textbf{SE}^{t-1}_{xg}$ with Eq. (6-7).\\
    \quad \textbf{Do Fuse Attentive Gating:}\\
    \quad $\overline{\textbf{h}^{t-1}_{xl,pre}}$ = \textbf{GATE} ($\textbf{h}^{t-1}_{xl,pre}$,$\textbf{h}^{t-1}_{xg,pre}$,$\textbf{h}^{t-1}_{xg,neig}$, $\textbf{SE}^{t-1}_{xg}$, $\textbf{SE}^{t-1}_{xl}$) \\
    \quad \textbf{On local graph $\mathcal{G}^t_X$ do:} \\
    \quad\quad Update $V_{pre}$'s embeddings to new state $\overline{\textbf{h}^{t-1}_{xl,pre}}$\\
    \quad\quad Add a directed edge from $v^X_{i-1}$ to $v_{i}$. \\
    \quad\quad Update nodes on graph $\mathcal{G}^t_X$ with Eq. (1-5).\\
    \quad \textbf{On global graph $\mathcal{G}^t_M$ do:} \\
    \quad\quad Add a directed edge from $v^M_{i-1}$ to $v_{i}$. \\
    \quad\quad Update nodes on graph $\mathcal{G}^t_M$ with Eq. (1-5).\\

\textbf{end for}\\
Get sequence embeddings $\textbf{SE}_{al,S_A}$, $\textbf{SE}_{bl,S_B}$ with Eq. (6-7).\\
\textbf{return} $\textbf{H}_{al}$, $\textbf{H}_{bl}$; $\textbf{SE}_{al,S_A}$, $\textbf{SE}_{bl,S_B}$\\
}
\end{algorithm} 


%
After $K$ steps of updating, we can obtain the embeddings of all nodes in $\mathcal{G}^t_A$ as $\textbf{H}^t_{al} = \{\textbf{h}^t_{al,1},\textbf{h}^t_{al,2},...,\textbf{h}^t_{al,i}\}$.
Then we adopt a strategy which attentively combines long-term and short-term preferences into the final sequential representation.
As for long-term preference, we apply the soft-attention mechanism to measure the varying importance of previous items and then aggregate them as a whole:
\begin{equation}
\begin{aligned}
\alpha^t_{al,k} &= p^\top_{al,t}\sigma(\mathbf{W}^t_{a1}\textbf{h}^t_{al,i}+\mathbf{W}^t_{a2}\textbf{h}^t_{al,k}+\mathbf{c}^t),\\
\textbf{SE}^t_{ac} &= \sum_{k=1}^i\alpha^t_{a,k}\textbf{h}^t_{al,k},
\end{aligned}
\end{equation}
where parameters $p^\top_{a,t} \in \mathbb{R}^D$ and $\mathbf{W}^t_{a1},\mathbf{W}^t_{a2} \in \mathbb{R}^{D \times D}$ control the weights of item embedding vectors.
As for \textit{short-term preference}, we concatenate the last item embedding which represents the current interest of the user with the above sequence embedding after aggregation and then take a linear transformation over them to generate the final sequence embedding:

\begin{equation}
\begin{aligned}
\textbf{SE}^t_{al} = \mathbf{W}^t_{a3}[\textbf{SE}^t_{ac};\textbf{h}^t_{al,i}].
\end{aligned}
\end{equation}

\section{dual dynamic graph modeling}\label{sec:alg}

We represent the thorough algorithm of the dual dynamic graph modeling module in Algorithm~\ref{alg:dual-dynamic-graph-algo}.
The input of this module includes single-domain and cross-domain behavior sequences, and the output consists of item embeddings and sequence embeddings.
The whole procedure of dual dynamic graph modeling can be divided into two parts, i.e., 1) dual dynamic graphs and 2) fuse attentive gate.
We describe how to construct dual dynamic graphs in line 14-20, and introduce how to transfer cross-domain information by fuse attentive gating mechanism in line 5-13.

\section{Experimental results}\label{sec:exp}

Here, we additionally report the experimental results on two Amazon datasets when
the cut-off of the ranked list is 10.
The results show that: (1) \modelname~outperforms all the baselines of SR, CDR, and CDSR.
(2) \modelname~also shows the superiority  over its variants, i.e., \modelname-L, \modelname-G, and \modelname-GA, indicating the effectiveness of each component in the dual dynamic graph modeling module.

\end{document}